\documentclass[proceedings]{JHEP3}
\usepackage{amsfonts}
\usepackage{amsmath}
\usepackage{epsfig}

\newbox\mybox

\newcommand\fverb{\setbox\mybox=\hbox\bgroup\verb}
\newcommand\fverbdo{\egroup\medskip\noindent\fbox{\unhbox\mybox}\ }
\newcommand\fverbit{\egroup\item[\fbox{\unhbox\mybox}]}
\conference{Time-independent approximations for time-dependent optical potentials}
\abstract{We explore the possibility of modifying the Lewis-Riesenfeld method of invariants developed originally to find exact solutions for time-dependent quantum mechanical systems
for the situation in which an exact invariant can be constructed, but the subsequently resulting
time-independent eigenvalue system is not solvable exactly. We propose to carry out this step in an approximate fashion, such as employing standard time-independent perturbation theory or the WKB approximation, 
and subsequently feeding the resulting approximated expressions back into the time-dependent scheme. We illustrate the quality of this approach by contrasting an exactly solvable solution to one obtained with
a perturbatively carried out second step for two types of explicitly time-dependent optical potentials.}

\title{Time-independent approximations for time-dependent optical potentials}
\author{Andreas Fring and Rebecca Tenney \\
Department of Mathematics, City University London,\\
Northampton Square, London EC1V 0HB, UK\\
E-mail: a.fring@city.ac.uk, rebecca.tenney@city.ac.uk}

\input{tcilatex}
\begin{document}

\section{Introduction}

Since Ashkin's discovery of the fact that radiation pressure from continuous
lasers can be used to trap small particles, almost fifty years ago \cite%
{ashkin1970}, various types of optical traps have been designed \cite%
{neuman2004optical}. Different variants of optical traps have found
widespead applications to fixate particles \cite{ashkin1986obs}, atoms \cite%
{metcalf2007laser}, molecules \cite{cronin2009optics}, and even living cells
such as viruses and bacteria \cite{ashkin1987opt,ashkin1987optical}. Once
the objects are fixed in the potential their properties can be investigated
in a very controlled and otherwise inaccessible manner. The general
underlying fundamental quantum mechanical description is provided by the
time-dependent Schr\"{o}dinger equation (TDSE) involving explicitly
time-dependent Hamiltonians in which the optical potentials typically take
on the generic form $V(x,t)=\kappa (t)V(x)$, with $V(x)$ describing the
trapping shape and $\kappa (t)$ some time-dependent modulation, e.g. \cite%
{o2001scaling}. There exist, however, also optical potentials that can not
be factorised in this manner, such as for instance periodic optical lattices 
\cite{fulton2006controlling}.

In general, there are only very few exact solutions known to the TDSE and
for many physical applications one relies predominatly on approximation
methods. Most approximation methods for time-dependent systems are carried
out on the level of the time-evolution operator, such as the adiabatic and
the sudden approximation \cite{born1928beweis}. In more concrete settings
less general approximation methods have been developed to account for the
specifics of the system. For instance, a very active field of research
involving time-dependent Hamiltonians is the area of strong laser fields 
\cite{amini2018symphony}. The systems considered in this context involve
Stark potentials of the form $V(x)+xE(t)$, with $E(t)$ describing a laser
field and the term $xE(t)$ dominating or being comparable in strength to the
potential $V(x)$. For these scenarios approximation schemes have been
developed as a mixture of perturbative expansions based on the Du-Hamel
formula carried out for time-evolution operator \cite{AC1,Rev1,AC3}. When
iterated, these expressions give rise to various versions of the Born
series. In the high intensity regime the two perturbative series, one in $%
V(x)$ and the other in $xE(t)$, are mixed and terminated after the first
iteration. The expressions obtained in this manner are commonly referred to
as the strong field approximation \cite%
{keldysh65,faisal73,reiss80,amini2018symphony}.

There are, however, methods purposefully designed to solve the TDSE exactly.
The Lewis-Riesenfeld method of invariants \cite{Lewis69} is one of them.
This scheme has been applied successfully to many models and scenarios, such
as the harmonic oscillator with time-dependent mass and frequency in one 
\cite{pedrosa1997exact} and two dimensions \cite%
{abdalla1988charged,bouguerra2006time}, the damped harmonic oscillator \cite%
{maamache2000exact}, a time-dependent Coulomb potential \cite%
{menouar2008exact}, a time-dependent Hamiltonians given in terms of linear
combinations of SU(1,1) and SU(2) generators \cite%
{lai1996time,maamache1998unitary,choi20071}, in the inverse construction of
time-dependent Hamiltonian \cite{chen2011lewis,fasihi2012non}, for systems
on noncommutative spaces in time-dependent backgrounds \cite%
{PhysRevD.90.084005}, time-dependent non-Hermitian Hamiltonian systems \cite%
{AndTom1,AndTom4,AndTom5,cen2019time} and other specific systems.

Even though many exact solutions were found, in comparison they are still
rare and for more involved potentials the method appears to be
nonapplicable. This is partly due to the attempt to complete the Lewis and
Riesenfeld method in its entirety. Here we propose to exploit a particular
feature of the scheme and modify it when it can only be completed partially.
In essence the method consists of three steps, of which often only the first
step, that is the construction of the invariants, can be completed. Instead
of abandoning the scheme one should recognize that with the completion of
the first step one has achieved a major simplification and has transformed
the system from a time-dependent first order differential equation to a
time-independent eigenvalue equation. Usually one demands in the next step
the solvability for this time-independent system, which again is a rare
property. For many concrete optical potentials this seems to be entirely
unachievable. We explore here the possibility of not terminating the
procedure at this stage, but instead to use perturbative theory or the WKB
approximation and subsequently feed the approximated expressions back into
the Lewis-Riesenfeld approach. Naturally one may also use other
approximation methods at this point.

We study here two concrete examples of optical potentials for which an exact
method may be obtained and then compare it to the result for which the
second step of the Lewis and Riesenfeld method has only been carried out
perturbatively. Our manuscript is organised as follows: In section 2 we
briefly recall the Lewis-Riesenfeld method of invariants and describe our
proposal of altering the second step in the scheme. In section 3 and 4 we
investigate in detail two concrete classes of optical potentials. The
potentials are chosen in such a way that the full scheme can be completed,
hence allowing a comparison with the approximated approach. Our conclusion
and an outlook into future problems is provided in section 5.

\section{An approximate Lewis-Riesenfeld method of invariants}

We start by recalling the key steps of the method of invariants and then
describe how they can be modified in an appropriate fashion. The scheme was
introduced originally by Lewis and Riesenfeld \cite{Lewis69}, for the
purpose of solving the TDSE 
\begin{equation}
i\hbar \partial _{t}\left\vert \psi _{n}\right\rangle =H(t)\left\vert \psi
_{n}\right\rangle ,  \label{SCH}
\end{equation}%
for the time-dependent or dressed states $\left\vert \psi _{n}\right\rangle $
associated to the explicitly time-dependent Hamiltonian $H(t)$.

\noindent The Lewis-Riesenfeld method of invariants is made up of three main
stages: The initial step in this approach consists of constructing a
time-dependent invariant $I(t)$ from the evolution equation%
\begin{equation}
\frac{dI(t)}{dt}=\partial _{t}I(t)+\frac{1}{i\hbar }[I(t),H(t)]=0.
\label{Inv}
\end{equation}%
Often this step can be completed and an exact form for the invariant $I(t)$
can be found. In the next step one needs to solve the corresponding
eigenvalue system of the invariant $I(t)$%
\begin{equation}
I(t)\left\vert \phi _{n}\right\rangle =\lambda _{n}\left\vert \phi
_{n}\right\rangle ,  \label{eigen}
\end{equation}%
for time-independent eigenvalues $\lambda _{n}$ and for the time-dependent
states $\left\vert \phi _{n}\right\rangle $. Provided the Hamiltonian $H(t)$
is Hermitian also the invariant $I(t)$ is Hermitian and therefore the
eigenvalues $\lambda _{n}$ are guaranteed to be real. The virtue of this
equation, compared to the TDSE in (\ref{SCH}), is that one has reduced the
original evolutionary problem in form of a first order differential equation
to an eigenvalue equation in which $t$ simply plays the role of a parameter.
Hence one just needs to solve a time-independent eigenvalue problem. To
complete this step the system in (\ref{eigen}) needs to be solvable. It is
this requirement one can weaken and employ \textit{time-independent }%
approximation methods to complete step two.

\noindent The final third step relates the eigenstates in (\ref{eigen}) with
the complete solution of the TDSE. It was shown in \cite{Lewis69} that the
states%
\begin{equation}
\left\vert \psi _{n}\right\rangle =e^{i\alpha _{n}(t)}\left\vert \phi
_{n}\right\rangle  \label{ph}
\end{equation}%
satisfy the TDSE (\ref{SCH}) provided that the real function $\alpha (t)$ in
(\ref{ph}) obeys%
\begin{equation}
\frac{d\alpha (t)}{dt}=\frac{1}{\hbar }\left\langle \phi _{n}\right\vert
i\hbar \partial _{t}-H(t)\left\vert \phi _{n}\right\rangle .  \label{pht}
\end{equation}%
Since all the quantities on the right hand side of (\ref{pht}) have been
obtained in the previous steps, one is left with a simple integration in
time to determine the phase $\alpha (t)$. These key equations serve mainly
for reference purposes and we refer the reader to \cite{Lewis69} for more
details.

For many systems we might succeed in carrying out the first step in the
procedure and construct an explicit expression for the invariant $I(t)$.
However, the process stalls often in the second step and for most
Hamiltonians the eigenvalue equation for the invariants $I(t)$ in (\ref%
{eigen}) can not be solved exactly. In this case we can, however, use
approximation methods. For instance, when the potential can be separated
into two terms, with one being dominating the other in absolute value, we
can set up a standard time-independent perturbation theory. For this purpose
let us briefly recall the main formulae in this approach.

We are splitting the invariant as%
\begin{equation}
I(t)=I_{0}(t)+\epsilon I_{p}(t),
\end{equation}%
and consider the eigenvalue equation for the full invariant and the
unperturbed one sparately%
\begin{equation}
I(t)\left\vert \phi _{n}\right\rangle =\lambda _{n}\left\vert \phi
_{n}\right\rangle ,\qquad \text{and\qquad }I_{0}(t)\left\vert \phi
_{n}^{(0)}\right\rangle =\lambda _{n}^{(0)}\left\vert \phi
_{n}^{(0)}\right\rangle .  \label{I0}
\end{equation}%
Assuming that within the perturbation term a small parameter $\epsilon \ll 1$
can be identified, we expand the eigenvalues and the eigenfunctions of the
unperturbed invariant as%
\begin{equation}
\lambda _{n}=\lambda _{n}^{(0)}+\epsilon \lambda _{n}^{(1)}+\epsilon
^{2}\lambda _{n}^{(2)}+\mathcal{O}(\epsilon ^{3}),\quad \text{and\quad }%
\left\vert \phi _{n}\right\rangle =\left\vert \phi _{n}^{(0)}\right\rangle
+\epsilon \left\vert \phi _{n}^{(1)}\right\rangle +\epsilon ^{2}\left\vert
\phi _{n}^{(2)}\right\rangle +\mathcal{O}(\epsilon ^{3}),
\end{equation}%
with $\lambda _{n}^{(k)}=1/k!\left. d\lambda _{n}/d\epsilon ^{k}\right\vert
_{\epsilon =0}$, $\left\vert \phi _{n}^{(k)}\right\rangle =1/k!\left. d\phi
_{n}/d\epsilon ^{k}\right\vert _{\epsilon =0}$. The first order corrections
to the eigenvalues and eigenstates of the invariants are then computed in
the standard fashion to%
\begin{equation}
\lambda _{n}^{(1)}=\left\langle \phi _{n}^{(0)}\right\vert I_{p}\left\vert
\phi _{n}^{(0)}\right\rangle ,\quad \text{and\quad }\left\vert \phi
_{n}^{(1)}\right\rangle =\sum\nolimits_{k\neq n}\frac{\left\langle \phi
_{k}^{(0)}\right\vert I_{p}\left\vert \phi _{n}^{(0)}\right\rangle }{\lambda
_{n}^{(0)}-\lambda _{k}^{(0)}}\left\vert \phi _{k}^{(0)}\right\rangle ,
\label{L1}
\end{equation}%
respectively. For orthonormal functions $\phi _{n}$, we obtain further
constraints on the normalization of contributions in the series%
\begin{eqnarray}
1=\left\langle \phi _{n}\right. \left\vert \phi _{n}\right\rangle
=\left\langle \phi _{n}^{(0)}\right. \left\vert \phi _{n}^{(0)}\right\rangle
&&+\epsilon \left( \left\langle \phi _{n}^{(0)}\right. \left\vert \phi
_{n}^{(1)}\right\rangle +\left\langle \phi _{n}^{(1)}\right. \left\vert \phi
_{n}^{(0)}\right\rangle \right) \\
&&+\epsilon ^{2}\left( \left\langle \phi _{n}^{(2)}\right. \left\vert \phi
_{n}^{(1)}\right\rangle +\left\langle \phi _{n}^{(1)}\right. \left\vert \phi
_{n}^{(1)}\right\rangle +\left\langle \phi _{n}^{(0)}\right. \left\vert \phi
_{n}^{(2)}\right\rangle \right) +\ldots ~~~~
\end{eqnarray}%
Thus if the zero order wavefunction is normalized to $1=\left\langle \phi
_{n}^{(0)}\right. \left\vert \phi _{n}^{(0)}\right\rangle $, we require the
higher order wave functions to satisfy the additional constraints 
\begin{equation}
\sum\limits_{k=0}^{\ell }\left\langle \phi _{n}^{(\ell -k)}\right.
\left\vert \phi _{n}^{(k)}\right\rangle =0.
\end{equation}%
Next we can use these expressions to obtain an approximate solution to the
TDSE. Denoting $\left\vert \phi _{n}\right\rangle ^{(1)}:=\left\vert \phi
_{n}^{(0)}\right\rangle +\epsilon \left\vert \phi _{n}^{(1)}\right\rangle $
we obtain 
\begin{equation}
\left\vert \psi _{n}\right\rangle ^{(1)}=e^{i\alpha _{n}^{(1)}(t)}\left\vert
\phi _{n}\right\rangle ^{(1)},\quad \text{and\quad }\alpha ^{(1)}(t)=\frac{1%
}{\hbar }\int dt\,^{(1)}\!\left\langle \phi _{n}\right\vert i\hbar \partial
_{t}-H(t)\left\vert \phi _{n}\right\rangle ^{(1)}\text{.}  \label{phi1}
\end{equation}

Alternatively, we may also solve the eigenvalue equation (\ref{eigen}) by
using the WKB approximation $\left\vert \phi _{n}^{\text{WKB}}\right\rangle $
and compute the phase using that expression%
\begin{equation}
\left\vert \psi _{n}^{\text{WKB}}\right\rangle =e^{i\alpha _{n}^{\text{WKB}%
}(t)}\left\vert \phi _{n}^{\text{WKB}}\right\rangle ,\quad \text{and\quad }%
\alpha ^{\text{WKB}}(t)=\frac{1}{\hbar }\int dt\left\langle \phi _{n}^{\text{%
WKB}}\right\vert i\hbar \partial _{t}-H(t)\left\vert \phi _{n}^{\text{WKB}%
}\right\rangle .  \label{WKBa}
\end{equation}%
Assuming that invariant $I(t)$ can be cast into the same form as a
time-independent Hamiltonian, with a stndard kinetic energy term and a
potential $V(\xi )$, the WKB approximation to first order in $\hbar $, see
e.g. \cite{bender2013advanced}, for the eigenvalue equation (\ref{eigen})
reads 
\begin{equation}
\hat{\phi}^{\text{WKB}}(\xi )=\frac{A}{\sqrt{p(\xi )}}e^{\frac{i}{\hbar }%
\int^{\xi }p(z)dz}+\frac{B}{\sqrt{p(\xi )}}e^{-\frac{i}{\hbar }\int^{\xi
}p(z)zx}  \label{wave1}
\end{equation}%
in the classically allowed region, $\lambda >V(\xi )$ and 
\begin{equation}
\hat{\phi}^{\text{WKB}}(\xi )=\frac{C}{\sqrt{q(\xi )}}e^{\frac{1}{\hbar }%
\int^{\xi }q(z)dz}+\frac{D}{\sqrt{q(\xi )}}e^{-\frac{1}{\hbar }\int^{\xi
}q(z)dz}  \label{wave2}
\end{equation}%
in the classically forbidden region $\lambda <V(\xi )$, where 
\begin{equation}
p(\xi ):=\sqrt{2[\lambda -V(\xi )]}\quad \text{and\quad }q(\xi ):=\sqrt{%
2[V(\xi )-\lambda ]}.
\end{equation}%
The constants $A$, $B$, $C$, $D$ need to be determined by the appropriate
asymptotic WKB matching and the normalisation conditions.

\noindent Let us now apply this approximation scheme to some concrete
systems.

\section{Time-dependent potentials with a Stark term}

We first demonstrate how to solve the TDSE (\ref{SCH}) for the
one-dimensional Stark Hamiltonian involving a time-dependent potential $%
V(x,t)$ 
\begin{equation}
H(t)=\frac{p^{2}}{2m}+\frac{m\omega ^{2}}{2}x^{2}+V(x,t)+xE(t).  \label{H}
\end{equation}%
In order to cover optical potentials of the form $V(x,t)$ in our treatment,
we are slightly more general than in the standard Stark Hamiltonian where
the potential is just depending on $x$ and allow for an explicit
time-dependence in the potential $V(x,t)$ as well as in an electric or laser
field $E(t)$. At first we assume that the potential factorizes as $%
V(x,t)=\kappa (t)V(x)$. When the laser field term involving $E(t)$ dominates
the potential term and $\kappa (t)=$const several well known and successful
approaches have been developed. For instance, the strong field approximation
is a mixture of perturbative expansions based on the Du-Hamel formula
carried out on the level of the time-evolution operator \cite%
{keldysh65,faisal73,reiss80,amini2018symphony}.

In our poposed approach we assume that the first step in the Lewis and
Riesenfeld can be carried out and resort to an approximation in form of
perturbation thory in the second step.

\subsection{Construction of time-independent invariants}

In order to carry out the first step in the Lewis-Riesenfeld approach to
solve time-dependent systems we need to construct the invariant $I(t)$ by
solving equation (\ref{Inv}) for a given Hamiltonian, (\ref{H}) in our case.
For this purpose one usually makes an Ansatz by assuming the invariant to be
of a similar form as the Hamiltonian%
\begin{equation}
I(t)=\frac{1}{2}\left[ \alpha (t)p^{2}+\beta (t)V(x)+\gamma (t)x+\delta
(t)\{x,p\}+\varepsilon (t)x^{2}\right] .  \label{in}
\end{equation}%
In our case it involves five unknown time-dependent coefficient functions $%
\alpha (t)$, $\beta (t)$, $\gamma (t)$, $\delta (t)$ and $\varepsilon (t)$.
As ususal we denote the anti-commutator by $\{A,B\}:=AB+BA$. The
substitution of (\ref{in}) into (\ref{Inv}) then yields the following first
order coupled differential equations as constraints%
\begin{eqnarray}
\dot{\alpha} &=&-2\frac{\delta }{m},\quad \gamma =2m\alpha E,\quad \dot{%
\gamma}=2\delta E,\quad \dot{\delta}=m\alpha \omega ^{2}-\frac{\varepsilon }{%
m},\quad \dot{\varepsilon}=2m\delta \omega ^{2},\quad  \label{co1} \\
\beta &=&m\alpha \kappa ,\quad \dot{\beta}=\delta \kappa x(\ln V)_{x}.
\label{co2}
\end{eqnarray}%
Remarkably, despite being overdetermined this system can be solved
consistently. We note that the equations in (\ref{co1}) and (\ref{co2})
almost decouple entirely from each other, being only related by $\delta $.
We solve (\ref{co1}) first by parameterizing $\alpha (t)=\sigma ^{2}(t)$ and
integrating twice%
\begin{equation}
\alpha =\sigma ^{2},\quad \gamma =2m\sigma ^{2}E(t),\quad \delta =-m\sigma 
\dot{\sigma},\quad \varepsilon =m^{2}\dot{\sigma}^{2}+m^{2}\frac{\tau }{%
\sigma ^{2}}.  \label{ade}
\end{equation}%
The auxiliary quantity $\sigma $ has to satisfy the nonlinear Ermakov-Pinney
(EP) \cite{Ermakov,Pinney} equation 
\begin{equation}
\ddot{\sigma}+\omega ^{2}\sigma =\frac{\tau }{\sigma ^{3}},  \label{Erma}
\end{equation}%
and in addition the electric field has to be parameterised by the solution
of the EP-equation $\sigma $ as 
\begin{equation}
E(t)=\frac{c}{\sigma ^{3}}.  \label{E}
\end{equation}%
The constants $c,\tau \in \mathbb{R}$ result from the integrations. We take
here $\tau >0$. Using the expression for $\delta $ from (\ref{ade}), we may
now also solve the set of equations in (\ref{co2}), obtaining 
\begin{equation}
\beta _{p}=m\sigma ^{2}\kappa _{p},\qquad V_{p}=c_{p}x^{p},\qquad \kappa
_{p}=\frac{\tilde{c}_{p}}{\sigma ^{2+p}},  \label{kp}
\end{equation}%
with real integration constants $c_{p},\tilde{c}_{p}$ and $p\in \mathbb{R}$.
This means that we can not choose the electric field $E(t)$ in our
Hamiltonian and the potential $V(x,t)$ entirely a priori and independently
from each other. Notice that we may extend the analysis by allowing the
constants $c_{p},\tilde{c}_{p}$ to be complex, hence opening up the
treatment to include non-Hermitian $\mathcal{PT}$-symmetric Hamiltonians 
\cite{Bender:1998ke,Alirev,PTbook}.

First we notice that the only time-independent potential is obtained for $%
p=-2$, so that the potential part in $H(t)$ becomes the solvable
Goldman-Krivchenko potential. Crucially, the constraining equations
involving the potential (\ref{co2}) \ decouple from the remaining ones and
since these equations are linear we may solve for potentials that factorize
termwise when expanded, that is $V(x,t)=\sum\nolimits_{p}$ $\kappa
_{p}(t)V_{p}(x)$. For instance, for a time-dependent Gaussian potential of
the form%
\begin{equation*}
V_{\text{Gauss}}(x,t)=A(t)\left( e^{-\lambda (t)x^{2}}-1\right)
=\sum\limits_{n=1}^{\infty }\kappa _{n}(t)V_{n}(x),
\end{equation*}%
we obtain%
\begin{equation}
V_{2n}=x^{2n},\quad \kappa _{n}=\frac{(-1)^{n}}{n!}\frac{1}{\sigma ^{2+2n}},
\end{equation}%
where have to restrict $A(t)=\lambda (t)=\sigma ^{-2}$. For another widely
used potential, the soft Coulomb potential of the form 
\begin{equation*}
V_{\text{sCoulomb}}(x,t)=A(t)\frac{1}{\sqrt{x^{2}+k^{2}a^{2}(t)}}%
=\sum\limits_{n=1}^{\infty }\kappa _{n}(t)V_{n}(x),
\end{equation*}%
with $k$ taken to be a real constant, we obtain%
\begin{equation}
V_{2n}=x^{2n},\quad \kappa _{n}=\frac{(-1)^{n}(2n)!}{(2n)!!(2n)!!}\frac{1}{%
\sigma ^{2+2n}k^{1+2n}},
\end{equation}%
where we have to restrict $A(t)=1/a(t)=\sigma ^{-1}$.

As mentioned, besides the potential, also the electric field is not entirely
unconstrained as they are mutually related via the EP-function $\sigma $.
However, as we shall demonstrate the solutions of the EP-equation are such
that it will still allow for a large class of interesting fields, notably
periodic, to be investigated in an exactly solvable manner. It was found by
Pinney \cite{Pinney} that the solutions to (\ref{Erma}) are 
\begin{equation}
\sigma =\sqrt{u_{1}^{2}+\tau \frac{u_{2}^{2}}{W^{2}}},  \label{SolEP}
\end{equation}%
where $u_{1}$, $u_{2}$ are the two linearly independent solutions of the
equation%
\begin{equation}
\ddot{u}+\omega ^{2}u=0,  \label{EPl}
\end{equation}%
and $W=u_{1}\dot{u}_{2}-\dot{u}_{1}u_{2}$ is the corresponding Wronskian.
Thus taking the two solutions of (\ref{EPl}) to be $u_{1}=A\sin (\omega t)$
and $u_{2}=B\cos (\omega t)$ with $A$, $B\in \mathbb{R}$, the solution to
the EP-equation (\ref{SolEP}) acquires the form 
\begin{equation}
\sigma (t)=\frac{1}{\sqrt{2}A\omega }\sqrt{\tau +A^{4}\omega ^{2}+\left(
\tau -A^{4}\omega ^{2}\right) \cos (2\omega t)}.  \label{solsig}
\end{equation}%
The function $\sigma (t)$ is regular since $\tau >0$. Therefore the electric
field follows to be 
\begin{equation}
E(t)=\frac{2\sqrt{2}\omega ^{3}E_{0}}{\left[ \omega ^{2}+\tau +(\omega
^{2}-\tau )\cos ^{2}(\omega t)\right] ^{3/2}},
\end{equation}%
where we have chosen the constants $c=E_{0}$ and $A=\sqrt{\tau }/\omega $
such that $E(0)=E_{0}$. We note that $\omega =\sqrt{\tau }$ is a special
point at which $\sigma (t)\rightarrow 1$ and also the field becomes
time-independent $E(t)\rightarrow E_{0}$.

Assembling everything we have completed the first step in the
Lewis-Riesenfeld construction procedure. The invariant acquires the form%
\begin{equation}
I(t)=\frac{\sigma ^{2}}{2}p^{2}+\frac{m^{2}}{2}\left( \dot{\sigma}^{2}+\frac{%
\tau }{\sigma ^{2}}\right) x^{2}+m\sum\nolimits_{p}c_{p}\tilde{c}_{p}\left( 
\frac{x}{\sigma }\right) ^{p}-\frac{1}{2}m\sigma \dot{\sigma}\{x,p\}+m\sigma
^{2}E(t)x,  \label{Inva}
\end{equation}%
with $\sigma (t)$ given by (\ref{solsig}) and free constants $\tau $, $m$, $%
\omega $, $c_{p}$, $\tilde{c}_{p}$ and $E_{0}$.

The second step, that is to solve the eigenvalue equation (\ref{eigen}), can
not be carried out exactly for all invariants $I(t)$ of the form in (\ref%
{Inva}). We therefore resort to a perturbative approach as outlined in the
previous section.

\subsection{Testing the semi-exact solutions}

\subsubsection{Exact computation}

A good indication about the quality of the perturbation theory and the WKB
approximation layed out above can be obtained by comparing both
approximations to an exact expression. For most cases this is of course not
possible, but taking in (\ref{H}) the potential for instance to be $%
V(x,t)=\kappa (t)x^{2}\,$, $\kappa (t)=2c_{\kappa }/\sigma ^{4}$, we obtain
an exactly solvable system that can serve as a benchmark. In this case the
expression (\ref{Inva}) for the invariant simply becomes%
\begin{equation}
I(t)=\frac{1}{2}\left[ \alpha p^{2}+\left( 2\beta +\varepsilon \right)
x^{2}+\delta \{x,p\}+\gamma x\right] ,  \label{II}
\end{equation}%
with $\alpha $, $\beta $, $\gamma $, $\delta $, $\varepsilon $ as specified
in (\ref{ade}). The eigenvalue equation is simplified further when
eliminating the anticommutator term $\{x,p\}$ by means of a unitarity
tranformation $U=\exp (i\delta x^{2}/2\alpha )$ and the subsequent
introduction of the new variable $\xi :=x/\sigma $. We compute%
\begin{equation}
\hat{I}=UIU^{-1}=-\frac{1}{2}\partial _{\xi }^{2}+\left( \frac{\tau }{2}%
m^{2}+mc_{\kappa }\right) \xi ^{2}+mE_{0}\xi .  \label{Iha}
\end{equation}%
The eigenvalue equation for the transformed, and in this case
time-independent, invariant $\hat{I}\chi (\xi )=\lambda \chi (\xi )$ is then
solved by 
\begin{equation}
\chi (\xi )=c_{1}D_{\mu _{+}}\left[ \sqrt{2}m^{1/4}\frac{(E_{0}+2c_{\kappa
}\xi +m\tau \xi )}{(2c_{\kappa }+m\tau )^{3/4}}\right] +c_{2}D_{\mu _{-}}%
\left[ i\sqrt{2}m^{1/4}\frac{(E_{0}+2c_{\kappa }\xi +m\tau \xi )}{%
(2c_{\kappa }+m\tau )^{3/4}}\right] ,
\end{equation}%
where $\mu _{\pm }=\pm (E_{0}^{2}m+4c_{\kappa }\lambda )/\sqrt{m}(2c_{\kappa
}+m\tau )^{3/2}-1/2$ and $D_{\nu }(z)$ denotes the parabolic cylinder
function. Demanding that the eigenfunctions vanish asymptotically, i.e. $%
\lim_{\xi \rightarrow \pm \infty }$ $\chi (\xi )=0$, imposes the constraint $%
\mu _{\pm }=n\in \mathbb{N}_{0}$ and thus quantizes the eigenvalues $\lambda
\rightarrow \lambda _{n}$. We discard the solution related to $D_{\mu _{-}}$%
, as its corresponding eigenvalues are not bounded from below. Hence, we are
left with the eigenfuctions and eigenvalues%
\begin{equation}
\chi _{n}(\xi )=c_{1}D_{n}\left[ \sqrt{2}m^{1/4}\frac{(E_{0}+2c_{\kappa }\xi
+m\tau \xi )}{(2c_{\kappa }+m\tau )^{3/4}}\right] ,~~\lambda _{n}=\left( n+%
\frac{1}{2}\right) \sqrt{2mc_{\kappa }+m^{2}\tau }-\frac{mE_{0}^{2}}{%
4c_{\kappa }+2m\tau }.  \label{Xn}
\end{equation}%
The eigenvalues are indeed time-independent as we expect in the context of
the Lewis-Riesenfeld approach. Assembling the above and using the
orthonormality property of the parabolic cylinder function $%
\int\nolimits_{-\infty }^{\infty }D_{n}(x)D_{m}(x)dx=n!\sqrt{2\pi }\delta
_{nm}$, we obtain the normalized eigenfunction $\phi _{n}=U^{-1}\chi _{n}$ \ 
\begin{equation}
\phi _{n}(x)=N_{n}D_{n}\left[ a+bx\right] e^{im\dot{\sigma}x^{2}/2\sigma },
\end{equation}%
for the operator $I$ in (\ref{II}) with%
\begin{equation}
N_{n}=\frac{m^{1/8}(2c_{\kappa }+m\tau )^{1/8}}{\sqrt{\sigma n!\sqrt{\pi }}}%
,\qquad a=\frac{\sqrt{2}m^{1/4}E_{0}}{(2c_{\kappa }+m\tau )^{3/4}},\qquad b=%
\frac{\sqrt{2}m^{1/4}(2c_{\kappa }+m\tau )^{1/4}}{\sigma }.  \label{ab}
\end{equation}%
Finally we compute the phase $\alpha _{n}(t)$ in (\ref{ph}) by means of (\ref%
{pht}). The right hand side yields\footnote{%
We used here the integrals%
\begin{eqnarray}
\int\limits_{-\infty }^{\infty }x^{2n}D_{2s+\delta }(x)D_{2r+\bar{\delta}%
}(x) &=&(-1)^{s}2^{s+r-n+\frac{1+\delta -\bar{\delta}}{2})}\sqrt{\pi }\Gamma
\left( \frac{1}{2}+s+\delta \right) \Gamma \left( 2n+1+\delta \right)  \\
&&\times _{3}\tilde{F}_{2}\left( -s,n+1,n+\frac{1}{2}+\delta ;\frac{1}{2}%
+\delta ,n-r+1+\frac{\delta -\bar{\delta}}{2};1\right)   \notag
\end{eqnarray}%
for $n,s,r\in \mathbb{N}_{0}$ and $(\delta ,\bar{\delta})=(0,1),(0,0),(1,1)$%
. The function $_{3}\tilde{F}_{2}\left( a,b,c;d,f;z\right) $ is the
regularized hypergeometric function defined as%
\begin{equation}
_{3}\tilde{F}_{2}\left( a,b,c;d,f;z\right) =\frac{1}{\Gamma \left( d\right)
\Gamma \left( f\right) }\sum\limits_{k=0}^{\infty }\frac{%
(a)_{k}(b)_{k}(c)_{k}}{(d)_{k}(f)_{k}}\frac{z^{k}}{k!},
\end{equation}%
with $(a)_{k}=\Gamma \left( a+k\right) /\Gamma \left( a\right) $ denoting
the Pochhammer symbol.}%
\begin{equation}
\left\langle \phi _{n}\right\vert i\partial _{t}-H(t)\left\vert \phi
_{n}\right\rangle =-\frac{\lambda _{n}}{m\sigma ^{2}}
\end{equation}%
so that phase becomes 
\begin{equation}
\alpha _{n}(t)=-\frac{1}{m\sqrt{\tau }}\lambda _{n}\arctan \left[ \frac{%
\sqrt{\tau }\tan \left( \omega t\right) }{\omega }\right] .
\end{equation}%
We notice that for $\omega \rightarrow \sqrt{\tau }$ this simply reduces to $%
\alpha _{n}(t)\rightarrow -\lambda _{n}/m$ and the Hamiltonian becomes
time-independent, so that this choice simply describes the time-independent
Schr\"{o}dinger equation.

\subsubsection{Perturbative computation}

Next we treat the term $V_{p}(x,t)=\kappa (t)x^{2}$ in the Hamiltonian as a
perturbation, so that we may view the system as being in the strong field
approximation. Accordingly we split up the invariant (\ref{II}) as $%
I(t)=I_{0}(t)+\epsilon I_{p}(t)$ with 
\begin{equation}
I_{0}(t)=\frac{1}{2}\left[ \alpha p^{2}+\varepsilon x^{2}+\delta
\{x,p\}+\gamma x\right] ,\quad ~~I_{p}(t)=\frac{m}{\sigma ^{2}}x^{2},
\end{equation}%
and the small expansion parameter is identified as $\epsilon \equiv
c_{\kappa }$. First we compute the correction to the eigenvalue of the
invariant. Solving the eigenvalue equation (\ref{I0}) and computing the
expectation values in (\ref{L1}) we obtain%
\begin{equation}
\lambda _{n}^{(0)}=\left( n+\frac{1}{2}\right) m\sqrt{\tau }-\frac{E_{0}^{2}%
}{2\tau },\quad \text{and\quad }\lambda _{n}^{(1)}=\frac{1}{\sqrt{\tau }}%
\left( n+\frac{1}{2}\right) +\frac{E_{0}^{2}}{m\tau ^{2}}.
\end{equation}%
As we expect, $\lambda _{n}^{(0)}+$ $c_{\kappa }\lambda _{n}^{(1)}$ is
precisely $\lambda _{n}$ in (\ref{Xn}) expanded up to first order in $%
c_{\kappa }$. Next we use (\ref{L1}) to compute the corrections to the
wavefunctions. There are only four terms contributing in the infinite sum.
We compute%
\begin{eqnarray}
\left\vert \phi _{n}^{(1)}\right\rangle &=&\frac{1}{4m\tau }\left[ \sqrt{%
n(n-1)}\left\vert \phi _{n-2}^{(0)}\right\rangle -\sqrt{(n+1)(n+2)}%
\left\vert \phi _{n+2}^{(0)}\right\rangle \right] \\
&&+\frac{E_{0}\sqrt{2}}{m^{3/2}\tau ^{7/4}}\left[ \sqrt{n+1}\left\vert \phi
_{n+1}^{(0)}\right\rangle -\sqrt{n}\left\vert \phi _{n-1}^{(0)}\right\rangle %
\right] .  \notag
\end{eqnarray}%
Finally we evaluate the perturbed expression for the phase $\alpha
_{n}^{(1)}(t)$ using equation (\ref{pht}). Up to first order we find%
\begin{equation}
\alpha _{n}^{(1)}(t)=-\frac{\lambda _{n}^{(0)}+c_{\kappa }\lambda _{n}^{(1)}%
}{m\sqrt{\tau }}\arctan \left[ \frac{\sqrt{\tau }\tan \left( \omega t\right) 
}{\omega }\right] .
\end{equation}%
Notice that for $\omega \rightarrow \sqrt{\tau }$ this simply reduces to $%
\alpha _{n}^{(1)}(t)=-t(\lambda _{n}^{(0)}+c_{\kappa }\lambda _{n}^{(1)})/m$%
. We have now obtained the full perturbative solution to the TDSE as $%
\left\vert \psi _{n}\right\rangle ^{(1)}$ as defined in (\ref{phi1}).

\subsubsection{WKB computation}

We start by determining the classical turning points $\xi _{\pm }$ from the
condition $\lambda =V(\xi )$. We find%
\begin{equation}
\xi _{\pm }=-\frac{E_{0}m\pm \sqrt{m}\sqrt{mE_{0}^{2}+2\lambda (m\tau
+2c_{\kappa })}}{m(m\tau +2c_{\kappa })},  \label{turning}
\end{equation}%
so that the WKB quantisation condition 
\begin{equation}
\int_{\xi _{-}}^{\xi _{+}}\sqrt{2(\lambda _{n}-V(\xi ))}d\xi =\pi \hbar
\left( n+\frac{1}{2}\right)
\end{equation}%
yields the \emph{exact }time-independent eigenvalues%
\begin{equation}
\lambda _{n}=\left( n+\frac{1}{2}\right) \hbar \sqrt{m}(2c_{\kappa }+m\tau
)^{1/2}-\frac{mE_{0}^{2}}{2(2c_{\kappa }+m\tau )},
\end{equation}%
as found above in (\ref{Xn}). Next we specify WKB wavefunction further.
Keeping in the classically forbidden regions $\xi \in (-\infty ,\xi _{-})$
and $\xi \in (\xi _{+},\infty )$ only the asymptotically decaying parts in (%
\ref{wave1}) and (\ref{wave2}), the corresponding WKB wavefunction are 
\begin{equation}
\hat{\phi}_{-}(\xi )=\frac{C_{3}(-1)^{n}}{\sqrt{q(\xi )}}\exp \left[ -\frac{1%
}{\hbar }\int_{\xi }^{\xi _{-}}q(z)dz\right] ,
\end{equation}%
and%
\begin{equation}
\hat{\phi}_{+}(\xi )=\frac{C_{3}}{\sqrt{q(\xi )}}\exp \left[ -\frac{1}{\hbar 
}\int_{\xi _{+}}^{\xi }q(z)dz\right] ,
\end{equation}%
respectively. At this point $C_{3}$ is the only undetermined constant left.
Carrying out the appropriate WKB matching we obtain for the classically
allowed region $\xi \in (\xi _{-},\xi _{+})$ the wavefunction 
\begin{equation}
\hat{\phi}_{b}(\xi )=\frac{2C_{3}(-1)^{n}}{\sqrt{p(\xi )}}\cos \left[ \frac{1%
}{\hbar }\int_{\xi _{-}}^{\xi }p(z)dz-\frac{\pi }{4}\right] .
\end{equation}%
We may compute these expressions by using the explicit expressions for the
functions $q(\xi )$ and $p(\xi )$. To do so we use the same abbreviated
comnstants $a$ and $b$ as in (\ref{ab}) that convert the potential,
eigenvalues and turning points into more compact forms 
\begin{equation}
V(\xi )=\frac{b^{4}}{8}\xi ^{2}+a\frac{b^{3}}{4}\xi ,\quad \lambda _{n}=%
\frac{1}{2}b^{2}\left( n+\frac{1}{2}\right) \hbar -\frac{a^{2}b^{2}}{8}%
,\quad \xi _{\pm }=\frac{\pm \sqrt{2}\sqrt{2b^{2}n\hbar +b^{2}\hbar }-ab}{%
b^{2}}.
\end{equation}%
After a lengthy computation we obtain the WKB wavefunction in the different
regions as%
\begin{eqnarray}
\hat{\phi}_{\pm }(\xi ) &=&\frac{C_{3}(\pm 1)^{n}2^{\mp \frac{n}{2}+\frac{1}{%
2}\mp \frac{1}{4}}\left[ \pm \frac{\sqrt{b^{2}(2n+1)}}{b}\right] ^{\mp
\left( n+\frac{1}{2}\right) }\left[ a+b\xi +\sqrt{q(\xi )}\right] ^{\pm
\left( n+\frac{1}{2}\right) }e^{\mp \frac{1}{4}\sqrt{q(\xi )}(a+b\xi )}}{%
\sqrt[4]{b^{2}q(\xi )}},  \notag \\
\hat{\phi}_{b}(\xi ) &=&\frac{C_{3}(-1)^{n}\cos \left[ \left( n+\frac{1}{2}%
\right) \arctan \left( \frac{a+b\xi }{\sqrt{p(\xi )}}\right) +\frac{1}{4}%
\sqrt{p(\xi )}(a+b\xi )+\frac{\pi n}{2}\right] }{\sqrt[4]{b^{2}p(\xi )}}.
\end{eqnarray}%
The last remaining constant $C_{3}$ may be fixed by the normalization
condition. Converting from the $\hat{I}$ eigenvalue equation back to the $I$
eigenvalue equation with $\phi =U^{-1}\hat{\phi}$ and the variable $\xi $ to 
$x/\sigma $, the normalisation condition amounts to 
\begin{equation}
\int_{-\infty }^{x_{-}}\phi _{-}^{\ast }(x)\phi
_{-}(x)dx+\int_{x_{-}}^{x_{+}}\phi _{b}^{\ast }(x)\phi
_{b}(x)dx+\int_{x_{+}}^{\infty }\phi _{+}^{\ast }(x)\phi _{+}(x)dx=1.
\label{no}
\end{equation}%
Evaluating the integrals in (\ref{no}), we find the $n$ independent constant 
\begin{equation}
C_{3}\approx \frac{b}{2\sqrt{\pi \sigma (t)}}.
\end{equation}%
Having found the WKB eigenfunction $\phi ^{\text{WKB}}$, we can now compute
the integrant in (\ref{WKBa}) that yields the WKB approximated
Lewis-Riesenfeld phase 
\begin{equation}
\alpha _{n}^{\text{WKB}}(t)\approx -\frac{1}{m\sqrt{\tau }}\lambda
_{n}\arctan \left[ \frac{\sqrt{\tau }\tan (\omega t)}{\omega }\right] .
\end{equation}%
Thus we have now obtained a WKB approximated solution $\psi _{n}^{\text{WKB}%
} $ to the time-dependent Schr\"{o}dinger equation as specified in (\ref%
{WKBa}). Let us now compare these three solutions.

\subsubsection{WKB versus pertubation theory versus exact solution}

In order to obtain an idea about the quality of these approximations let us
compute some physical quantities in an exact and perturbative manner and
subequently compare them. For the exact case we find the expectation values
for the momentum, position and their squares as%
\begin{equation}
\begin{array}{ll}
\left\langle \psi _{n}\right\vert x\left\vert \psi _{n}\right\rangle =-\frac{%
E_{0}\sigma }{2c_{\kappa }+m\tau },~ & \left\langle \psi _{n}\right\vert
x^{2}\left\vert \psi _{n}\right\rangle =\frac{E_{0}^{2}\sigma ^{2}}{%
(2c_{\kappa }+m\tau )^{2}}+\frac{(2n+1)\sigma ^{2}}{2\sqrt{m}\sqrt{%
2c_{\kappa }+m\tau }}, \\ 
\left\langle \psi _{n}\right\vert p\left\vert \psi _{n}\right\rangle =-\frac{%
mE_{0}\dot{\sigma}}{2c_{\kappa }+m\tau }, & \left\langle \psi
_{n}\right\vert p^{2}\left\vert \psi _{n}\right\rangle =\frac{m^{2}E_{0}^{2}%
\dot{\sigma}^{2}}{(2c_{\kappa }+m\tau )^{2}}+\frac{(2n+1)m^{1/2}}{2\sigma
^{2}\sqrt{2c_{\kappa }+m\tau }}\left( 2c_{\kappa }+m\tau +m\sigma ^{2}\dot{%
\sigma}^{2}\right) ,%
\end{array}%
\end{equation}%
such that the uncertainty relation becomes%
\begin{equation}
\Delta x\Delta p=(n+\frac{1}{2})\sqrt{1+\frac{m(\tau -\omega ^{2})^{2}\sin
^{2}(2\omega t)}{4\omega ^{2}(2c_{\kappa }+m\tau )}},
\end{equation}%
where as usual the squared uncertainty is defined as the squared standard
deviation $\Delta A^{2}:=\left\langle \psi _{n}\right\vert A^{2}\left\vert
\psi _{n}\right\rangle -\left\langle \psi _{n}\right\vert A\left\vert \psi
_{n}\right\rangle ^{2}$ for $A=x,p$. Since the square root is always greater
or equal to $1$, the bound in the uncertainty relation $\Delta x\Delta p\geq
1/2$ is always respected.

From the perturbed solution $\left\vert \psi _{n}\right\rangle ^{(1)}$ we
find%
\begin{equation}
\begin{array}{l}
\left\langle \psi _{n}\right\vert x\left\vert \psi _{n}\right\rangle ^{(1)}=-%
\frac{E_{0}\sigma }{m\tau }+c_{k}\frac{2E_{0}\sigma }{m^{2}\tau ^{2}}, \\ 
\left\langle \psi _{n}\right\vert p\left\vert \psi _{n}\right\rangle ^{(1)}=-%
\frac{E_{0}\dot{\sigma}}{\tau }+c_{k}\frac{2E_{0}\dot{\sigma}}{m\tau ^{2}},
\\ 
\left\langle \psi _{n}\right\vert x^{2}\left\vert \psi _{n}\right\rangle
^{(1)}=\frac{2E_{0}^{2}+(2n+1)m\tau ^{3/2}}{2m^{2}\tau ^{2}}\sigma ^{2}-c_{k}%
\frac{8E_{0}^{2}+(2n+1)m\tau ^{3/2}}{2m^{3}\tau ^{3}}\sigma ^{2}, \\ 
\left\langle \psi _{n}\right\vert p^{2}\left\vert \psi _{n}\right\rangle
^{(1)}=\frac{E_{0}^{2}\dot{\sigma}^{2}}{\tau ^{2}}+\frac{m\sqrt{\tau }(2n+1)%
}{2\sigma ^{2}}+\frac{m(2n+1)\dot{\sigma}^{2}}{2\sqrt{\tau }}+c_{k}\left( 
\frac{n+1/2}{\sqrt{\tau }\sigma ^{2}}-\frac{4E_{0}^{2}\dot{\sigma}^{2}}{%
m\tau ^{3}}-\frac{(2n+1)\dot{\sigma}^{2}}{2\tau ^{3/2}}\right) .%
\end{array}%
\end{equation}%
and%
\begin{equation}
\Delta x\Delta p^{(1)}=\left( n+\frac{1}{2}\right) \left[ \sqrt{1+\frac{%
\sigma ^{2}\dot{\sigma}^{2}}{\tau }}-c_{k}\left( \frac{\sigma ^{2}\dot{\sigma%
}^{2}}{m\tau ^{3/2}\sqrt{\tau +\sigma ^{2}\dot{\sigma}^{2}}}\right) \right] .
\end{equation}%
These expresssions coincide with the exact expressions expanded up to order
one in $c_{k}$. In figure \ref{Fig1} we compare the time-dependent
expectation values for $x$, $x^{2}$, $p$, $p^{2}$ computed in an exact way
with those computed in a perturbative fashion. In general the agreement is
very good for small values of $c_{k}$. Overall the agreement is increasing
for large values of $n$ as well as $m$ and for $\omega $ approaching $\sqrt{%
\tau }$.

\FIGURE{ \epsfig{file=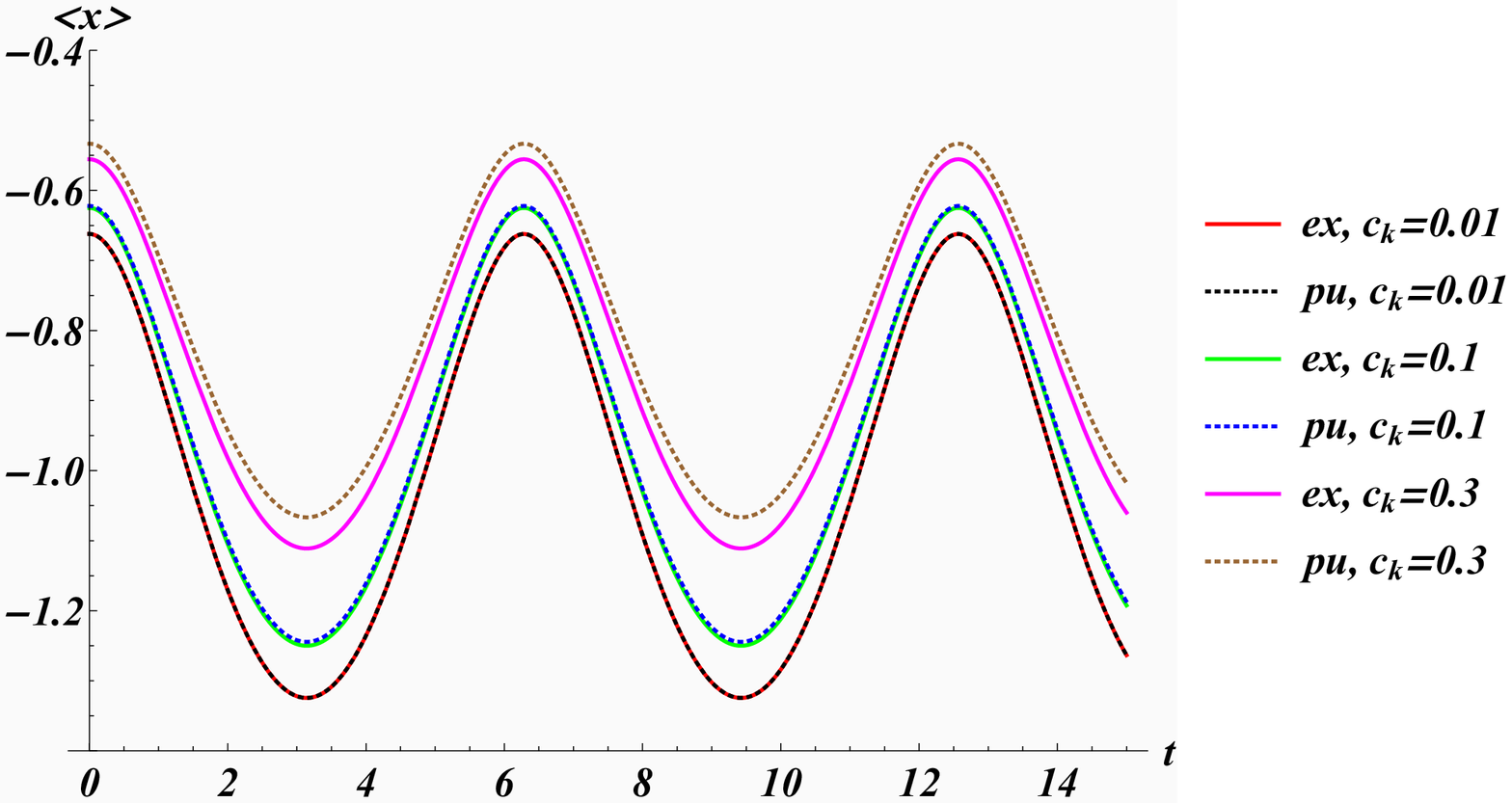,width=7.2cm} \epsfig{file=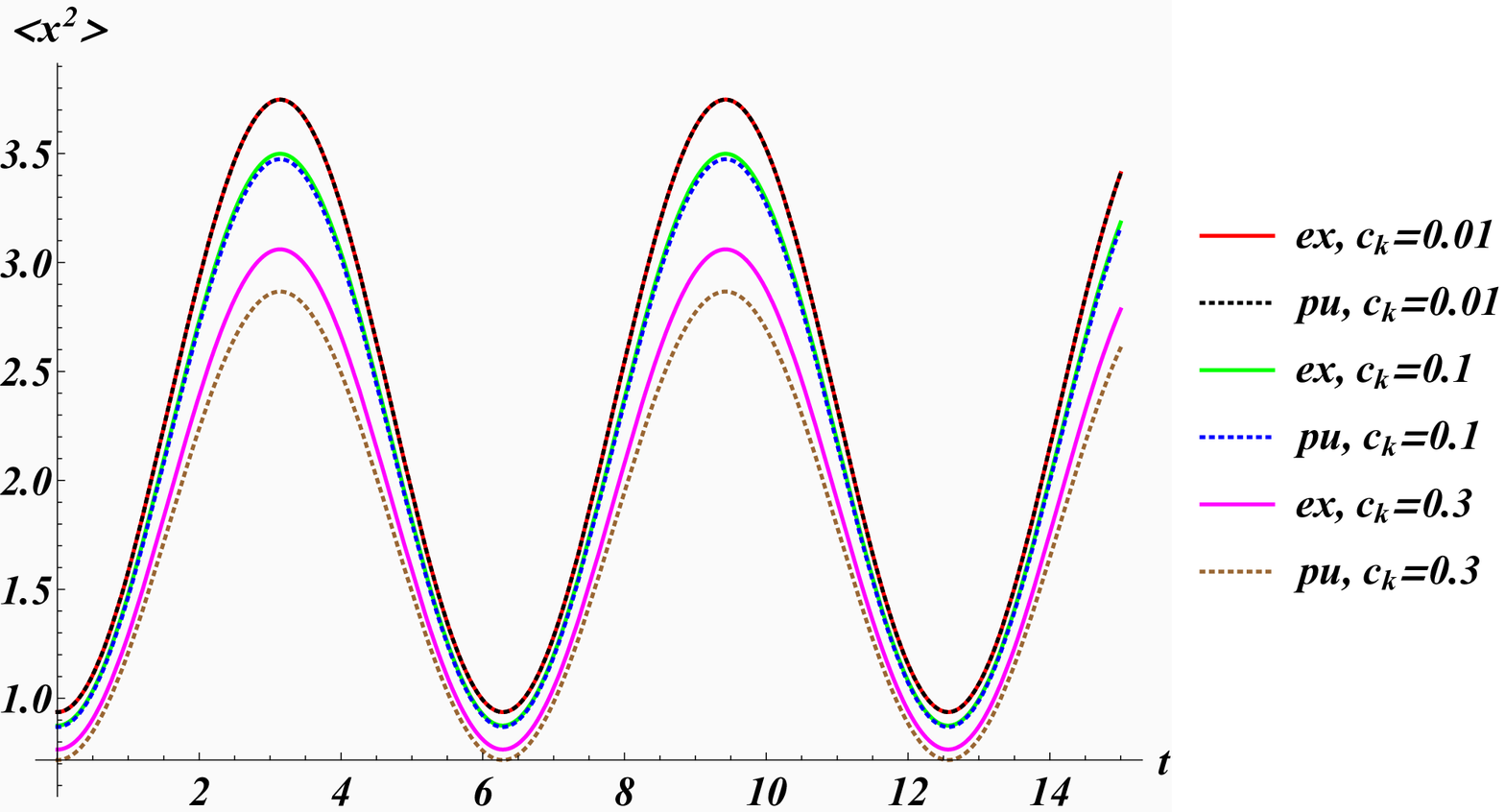,width=7.2cm} \epsfig{file=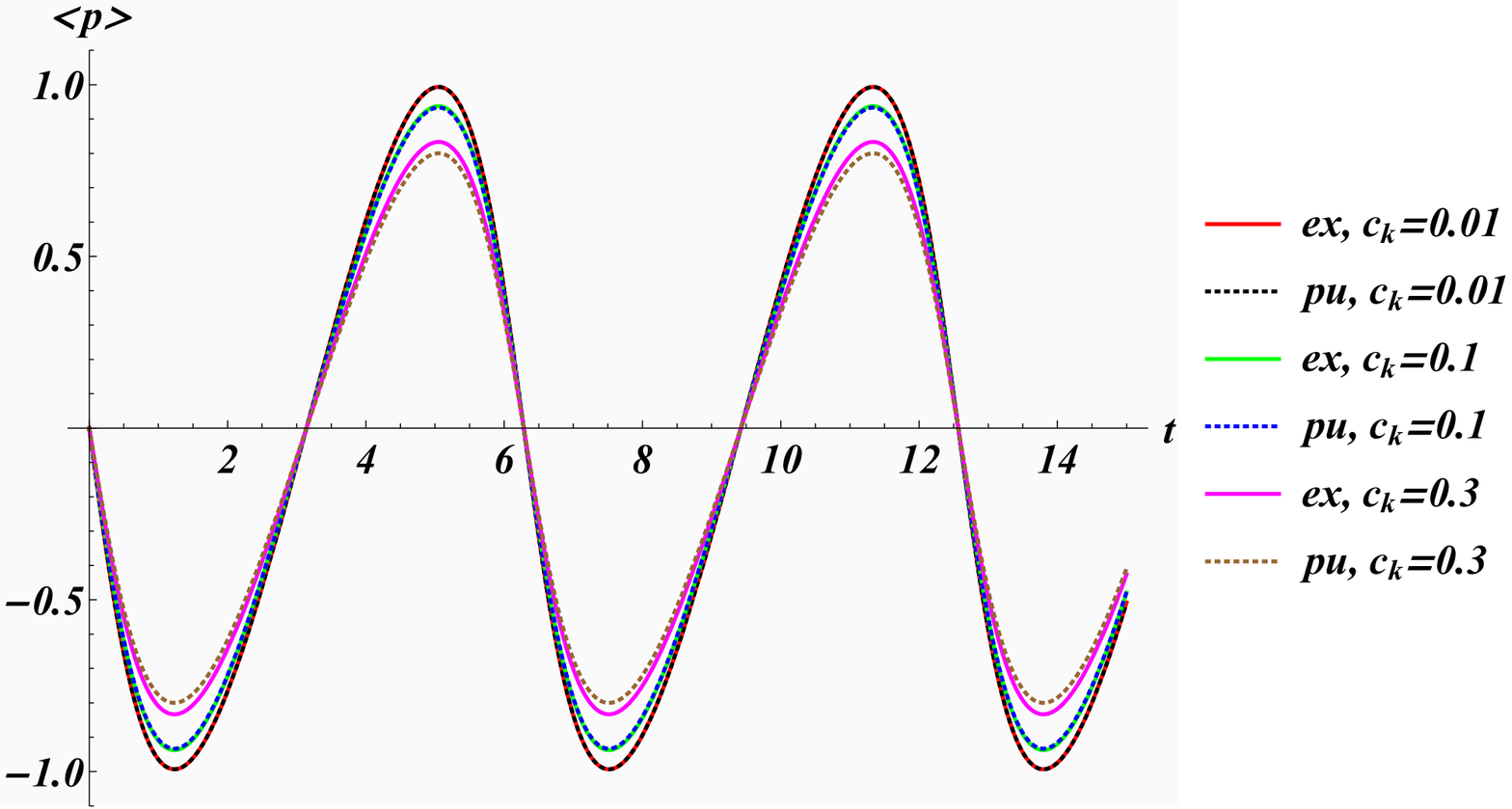,width=7.2cm} \epsfig{file=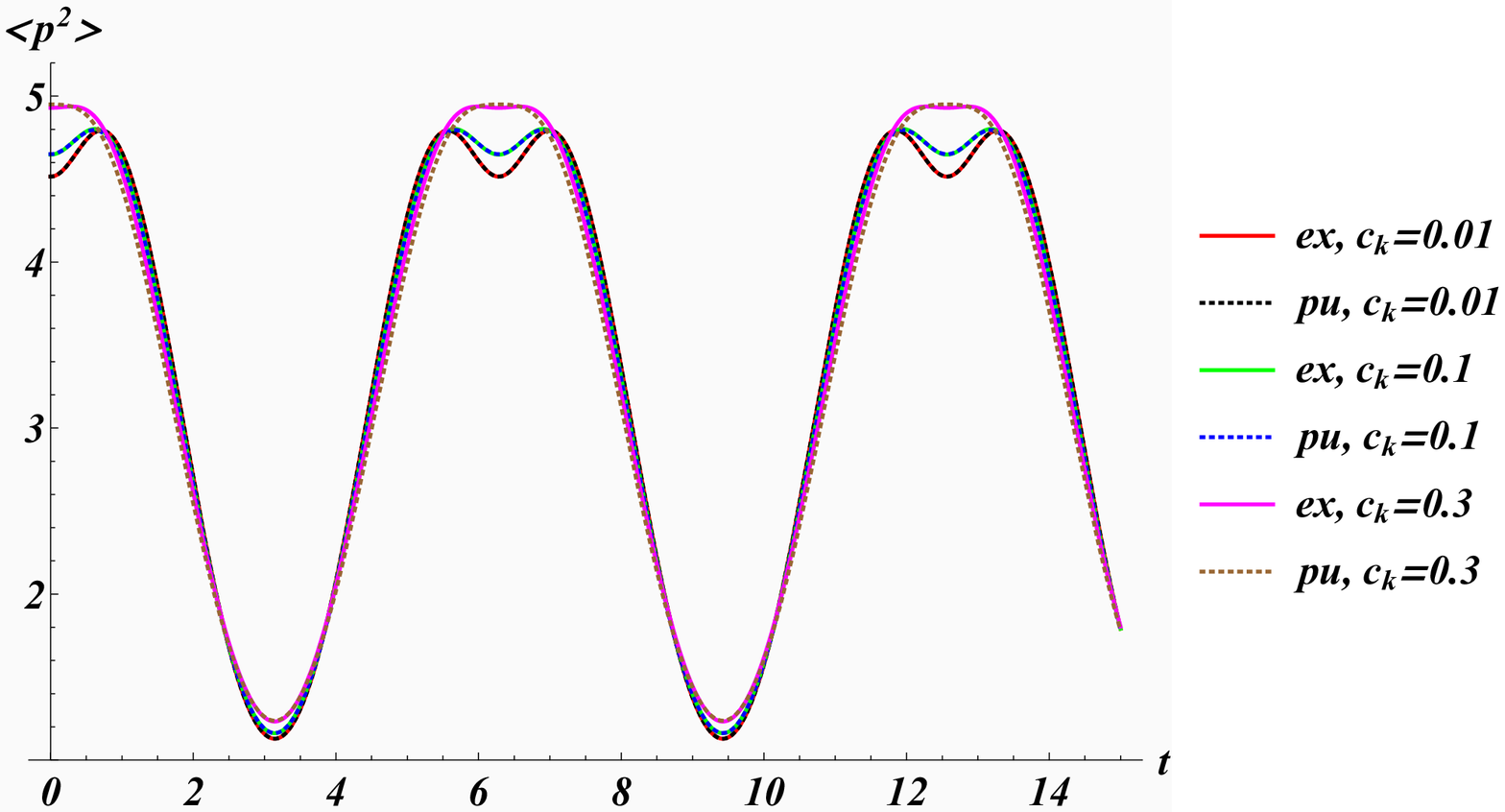,width=7.2cm}
        \caption{Exact versus perturbative expectation values for $x$, $x^{2}$, $p$, $p^{2}$ for $E_{0}=2$, $\omega =1/2$, $\tau =1$, $m=3$ and $n=1$ for different values of the expansion parameter $c_k$.}
        \label{Fig1}}

A further useful quantity to compute that illustrates the quality of the
perturbative approach is the autocorrelation function 
\begin{equation}
A_{n}(t):=\left\vert \left\langle \psi _{n}(t)\right. \left\vert \psi
_{n}(0)\right\rangle \right\vert .  \label{auto}
\end{equation}%
Unlike the expectation values for position, momenta and their squares the
autocorrelation function also captures the influence of the time-dependent
phase $\alpha (t)$. We depict this function in figure \ref{Fig2}. In this
case the overall agreement decreases for larger values of $n$.

\FIGURE{ \epsfig{file=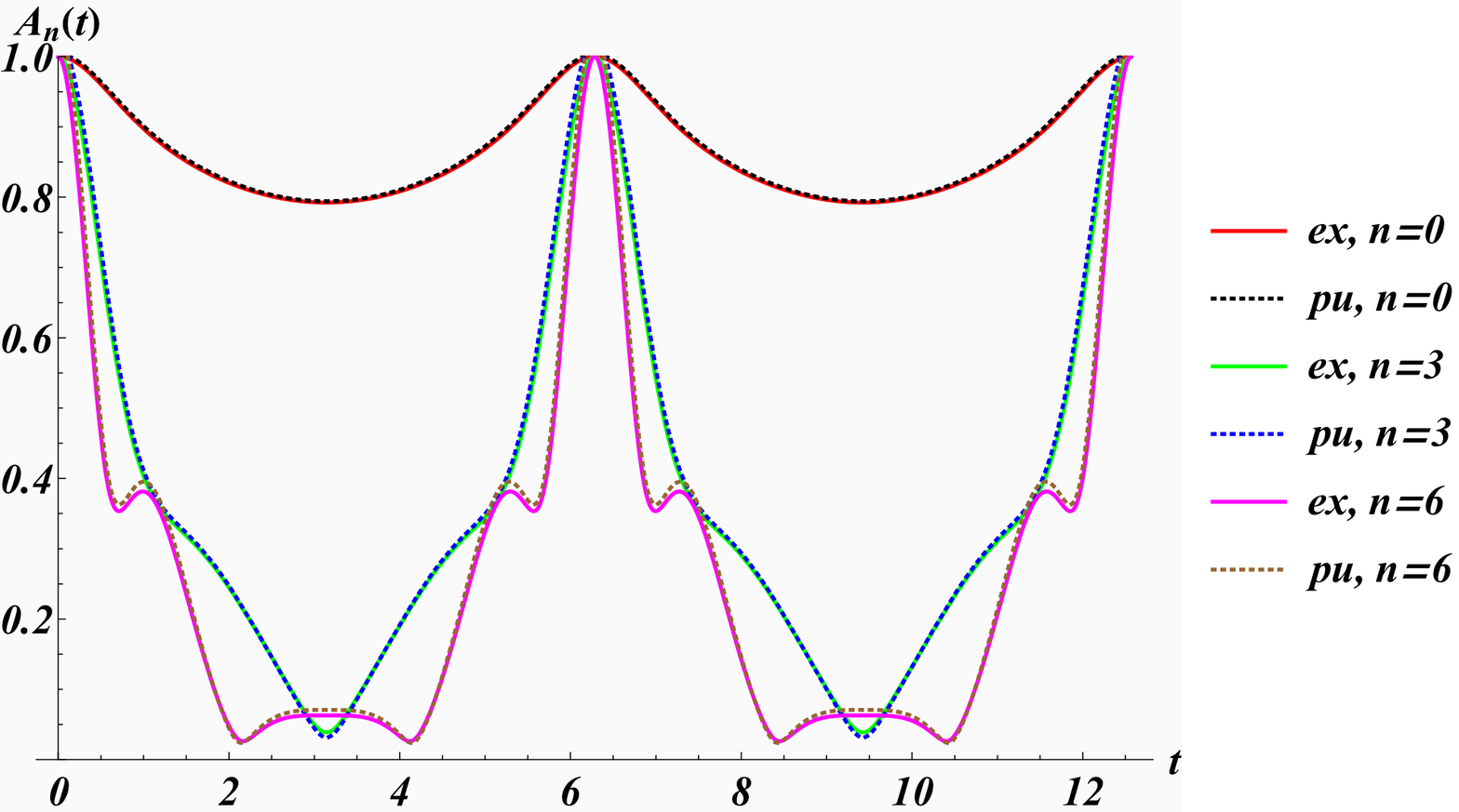,width=7.2cm} \epsfig{file=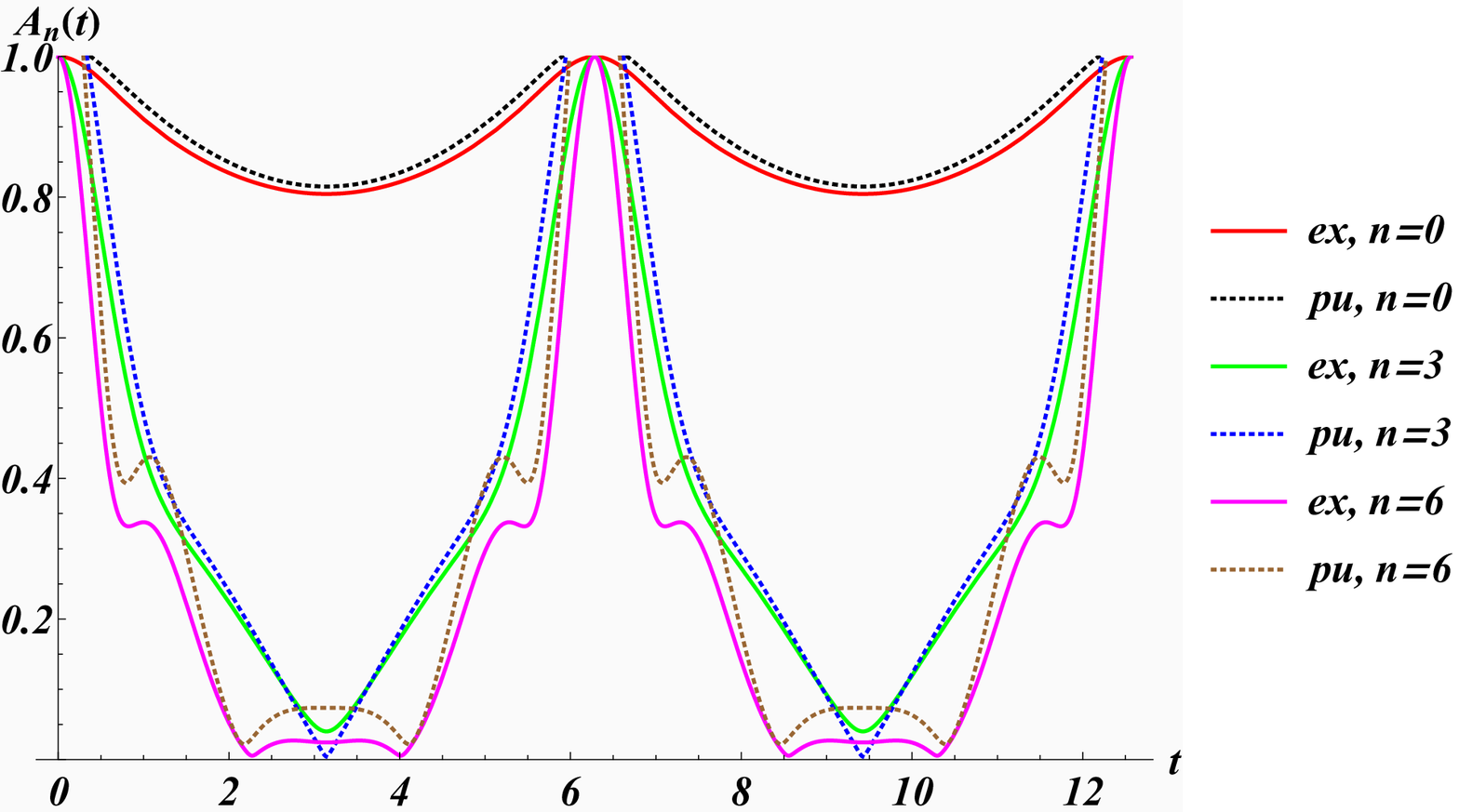,width=7.2cm}
        \caption{Exact versus perturbative autocorrelation function for $E_{0}=2$, $\omega =1/2$, $\tau =1$, $m=3$, different values 
         for $n$ with $c_k=0.1$ in the left panel and $c_k=0.3$ in the right panel.}
        \label{Fig2}}

Next we compare directly the wave functions obtained three alternative ways.
Figure \ref{Fig5} shows an extremely good agreement between the WKB
approximation and the exact solution, except near the turning points $\xi
_{\pm }$ where the WKB approximation is singular. The pertubative solution
is in very good agreement with the exact solution for small values of $%
c_{\kappa }$, as expected. With increasing values of $c_{\kappa }$ the
perturbative solution starts to deviate stronger in the negative regime for $%
\xi $ and large values of $n$.

\FIGURE{ \epsfig{file=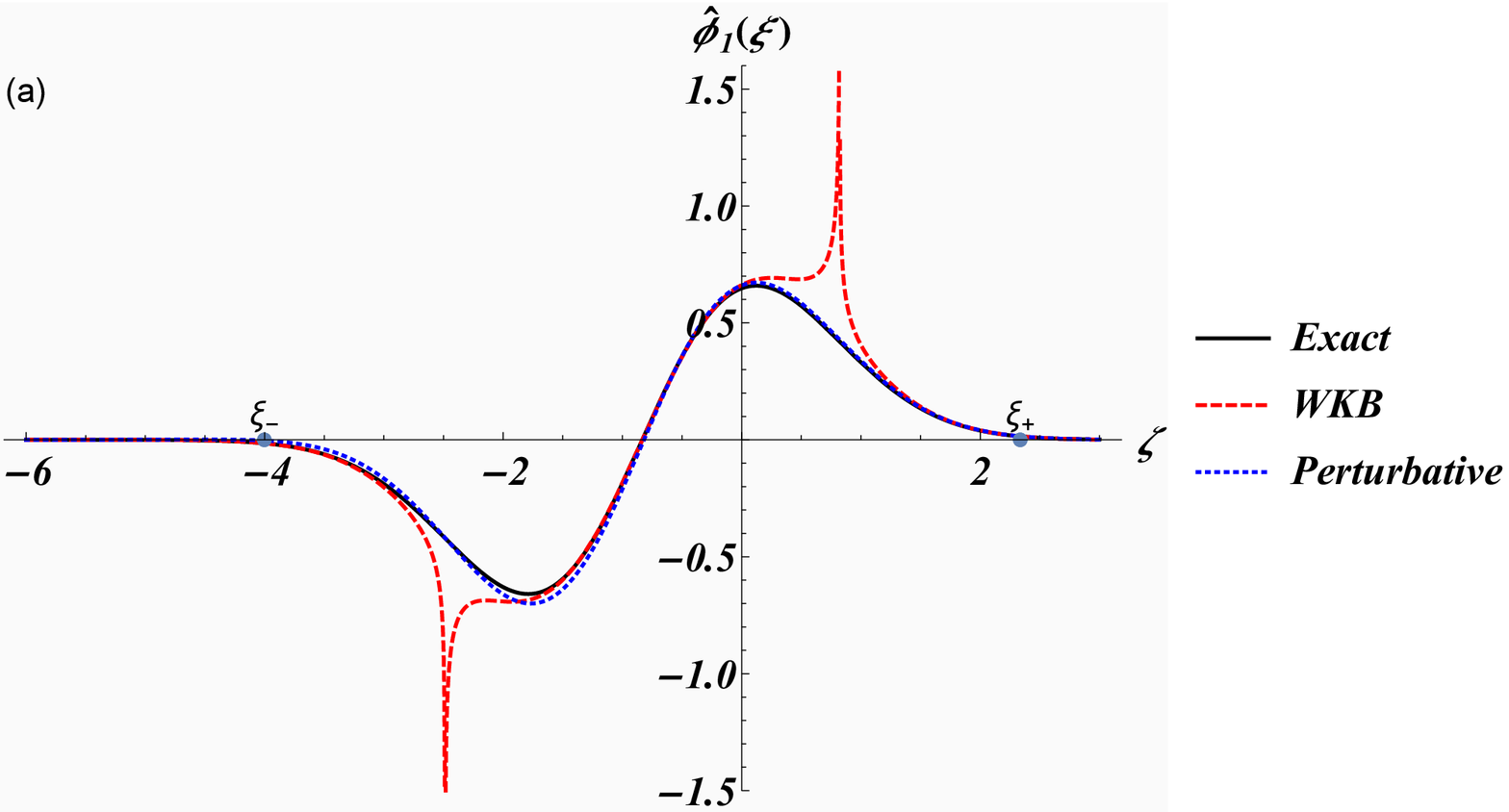, width=7.2cm} \epsfig{file=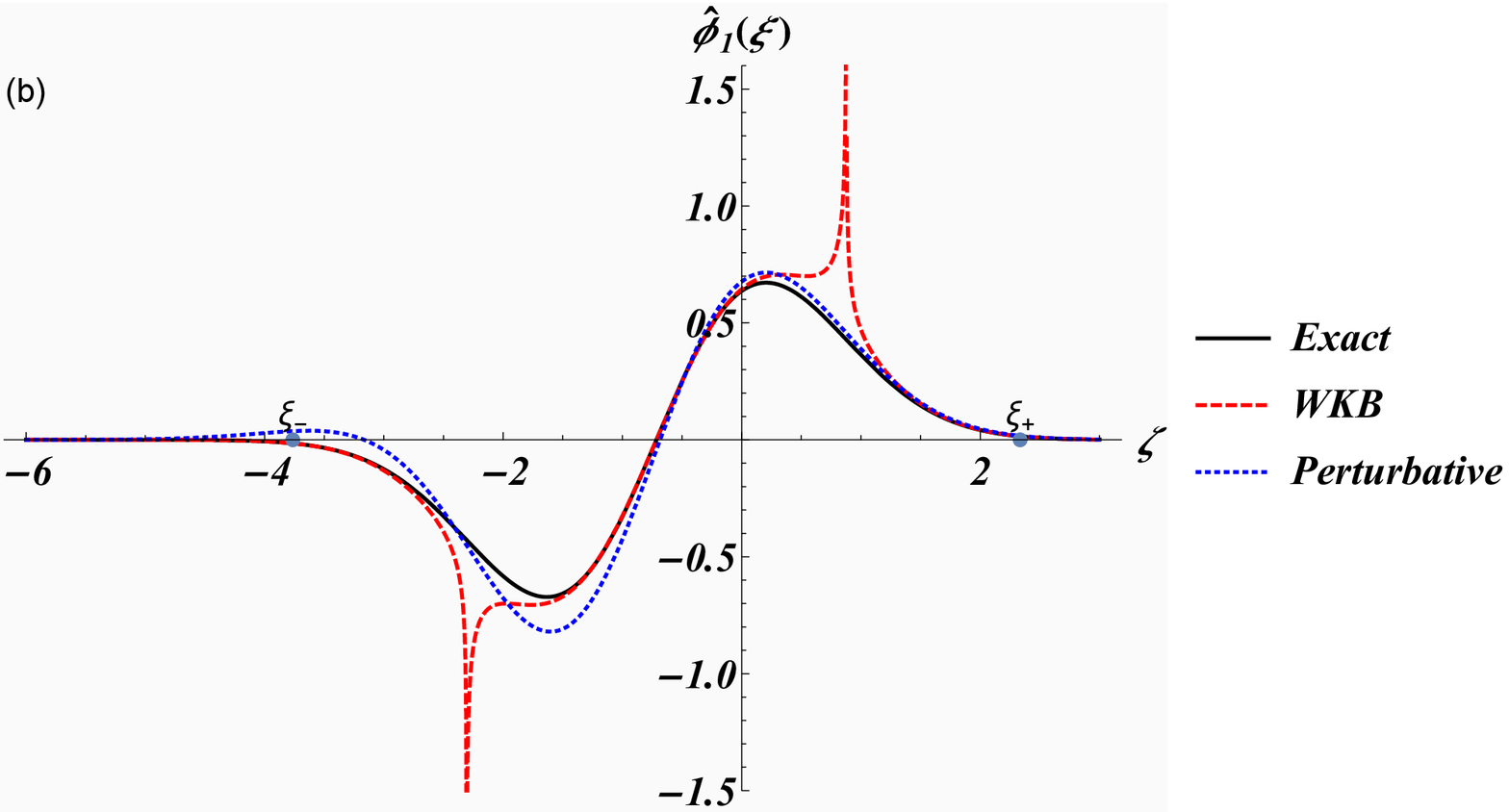, width=7.2cm} \epsfig{file=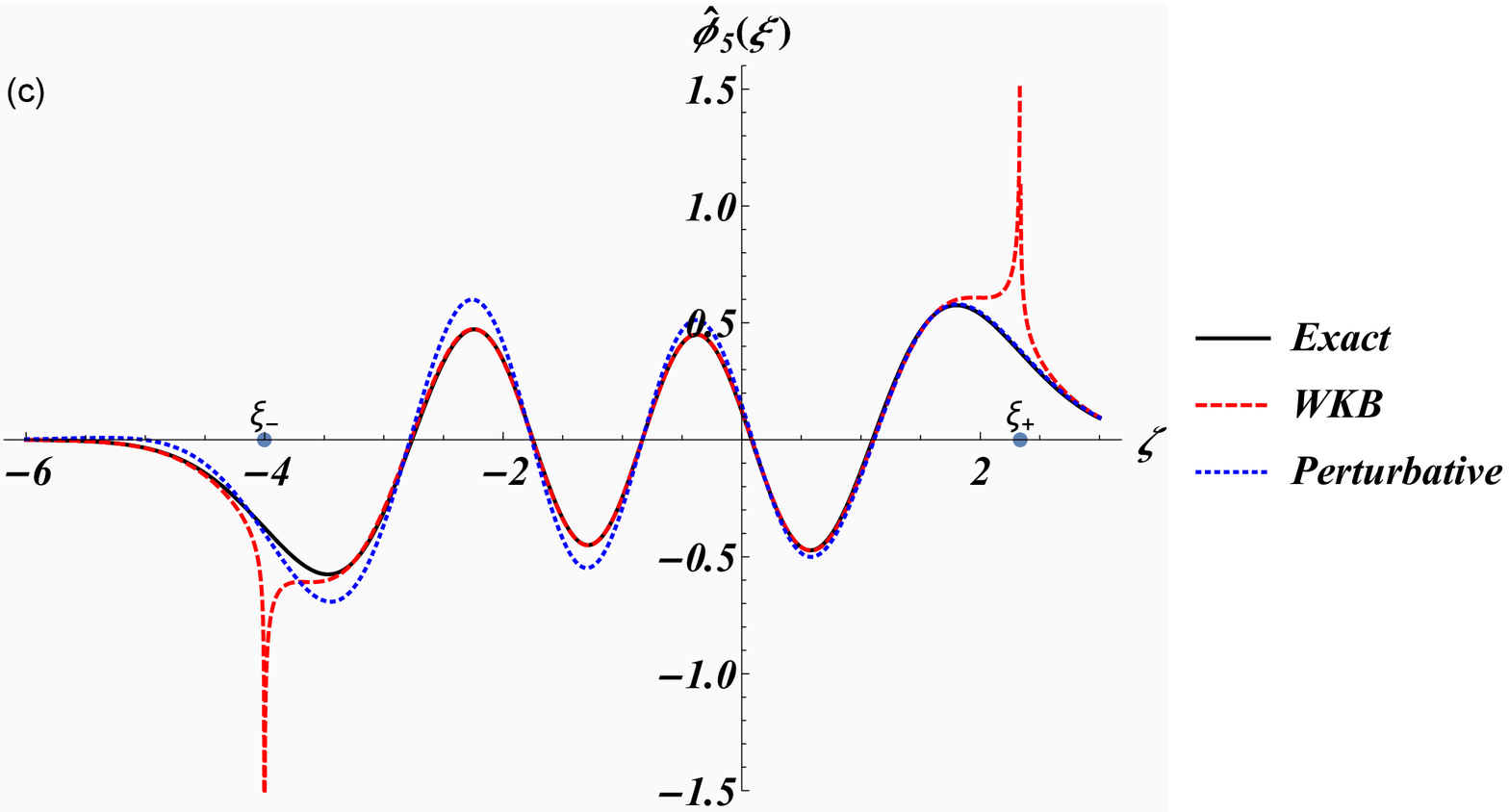, width=7.2cm} \epsfig{file=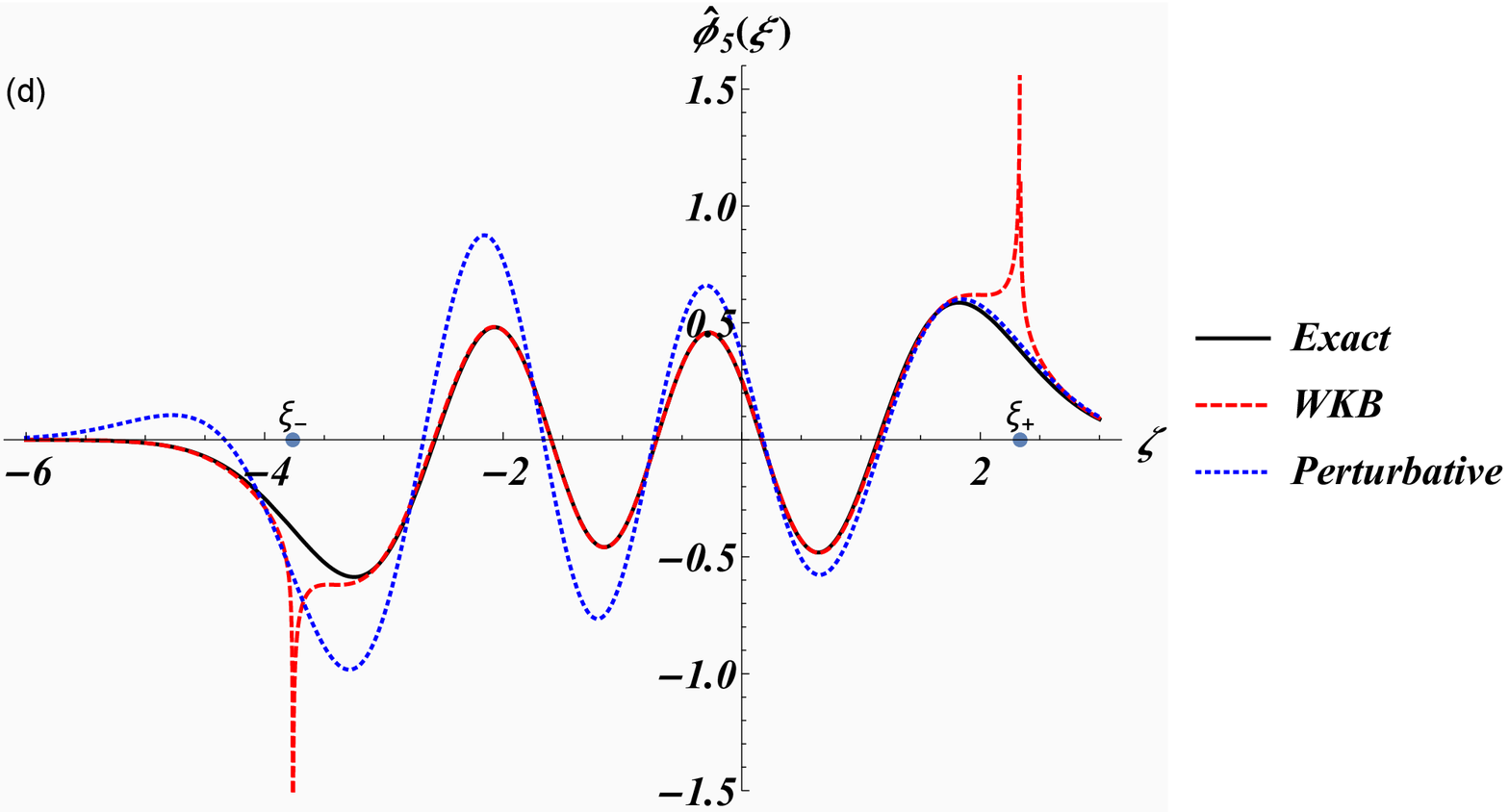, width=7.2cm}
\caption{Exact versus WKB and perturbative solutions to the time-independent
eigenvalue equation (\ref{eigen}) for the invariant $\hat{I}$ with $\hbar =1$, $E_{0}=1$, $m=1$, $n=1$, $\tau =1$ and $c_{\kappa }=0.18$ in panels (a), (c), $c_{\kappa
}=0.2$ in panels (b), (d). $\xi _{\pm }$ are the classical turning points (\ref{turning}). }
        \label{Fig5}}

Let us next see how these properties are inherited in the time-dependent
system. Figure \ref{Fig6} displays the real part of the full time-dependent
wave function. We observe the oscillation of the turning points with time
that enter through the function $\sigma (t)$. As in the time-independent
case, extremely good agreement between the WKB approximation and the exact
solution, except near the turning points $x_{\pm }$. The perturbative
solution slightly overshoots at the maxima and minima, especially in the
negative time regime. The discrepancy becomes worse for larger values of $n$%
, which we do not show here.

\FIGURE{ \epsfig{file=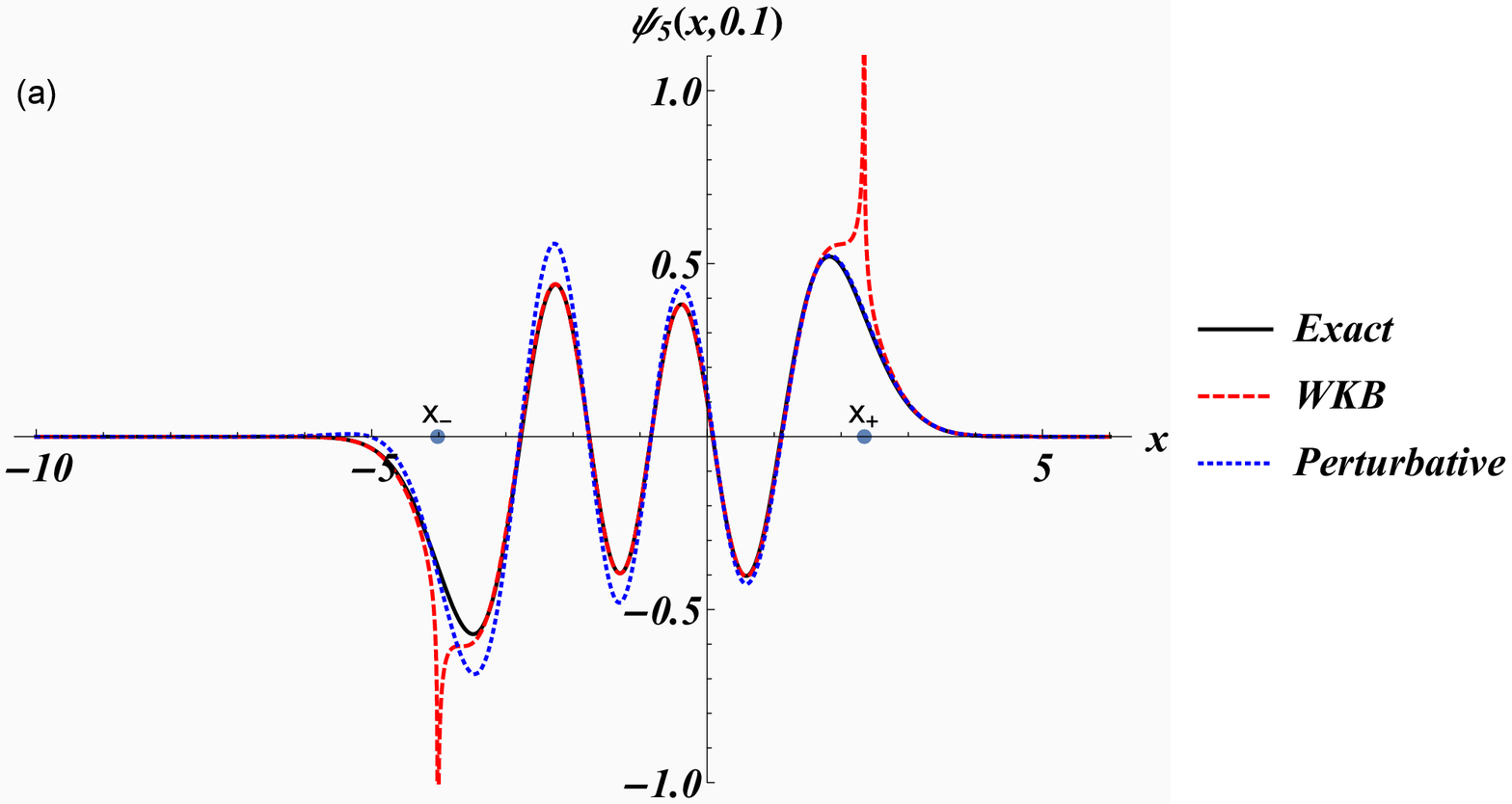, width=7.2cm} \epsfig{file=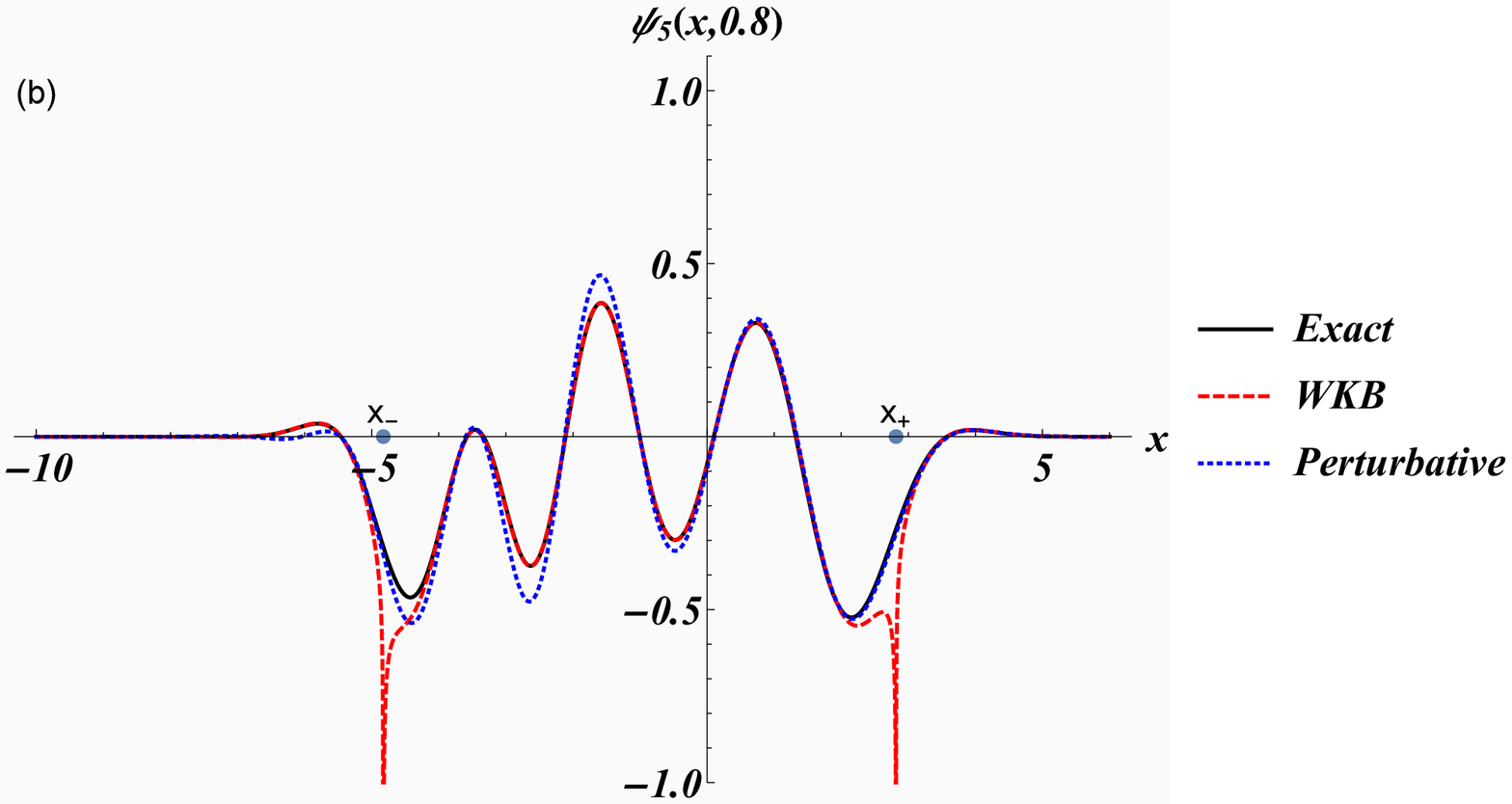, width=7.2cm} \epsfig{file=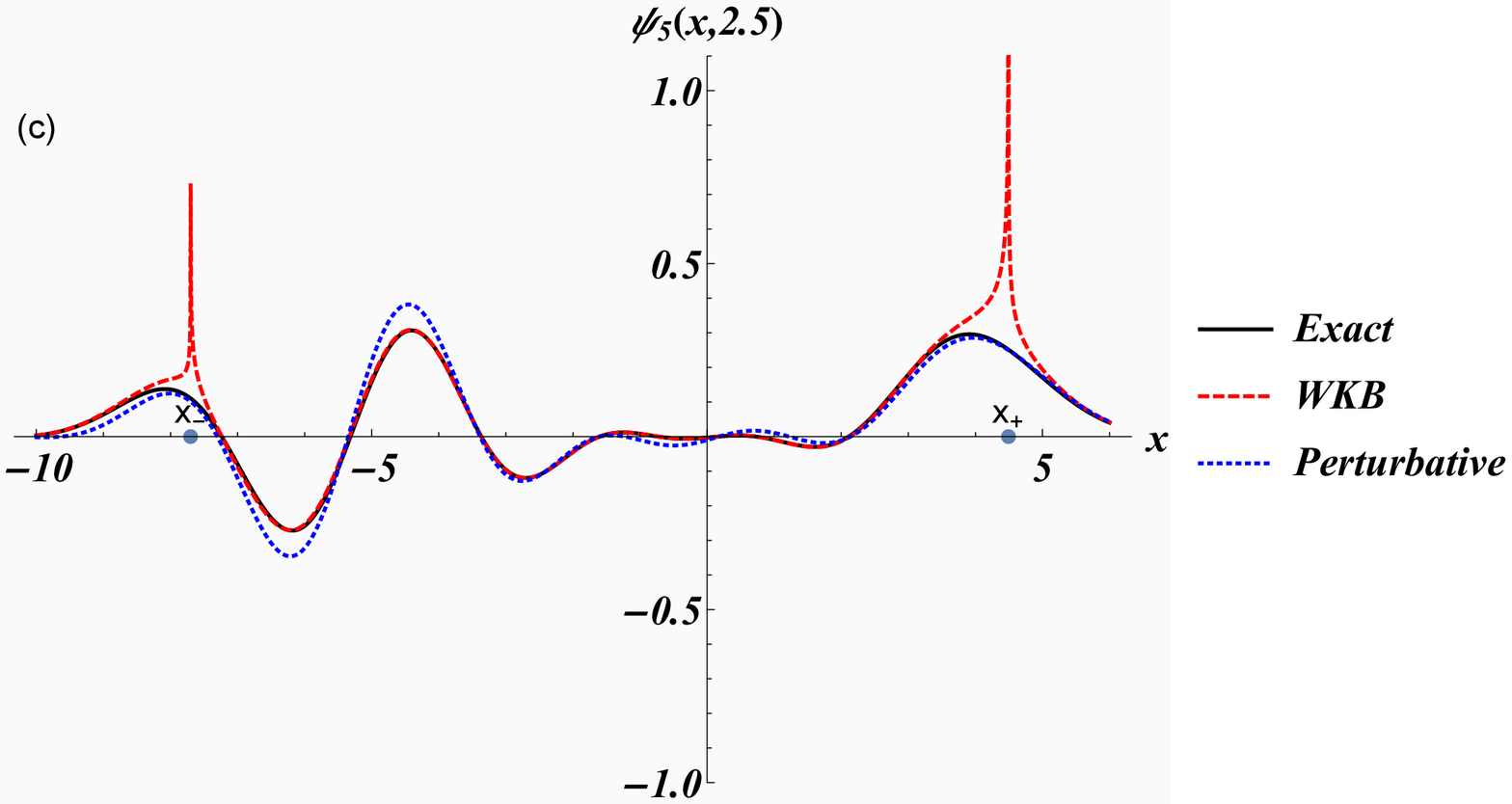, width=7.2cm} \epsfig{file=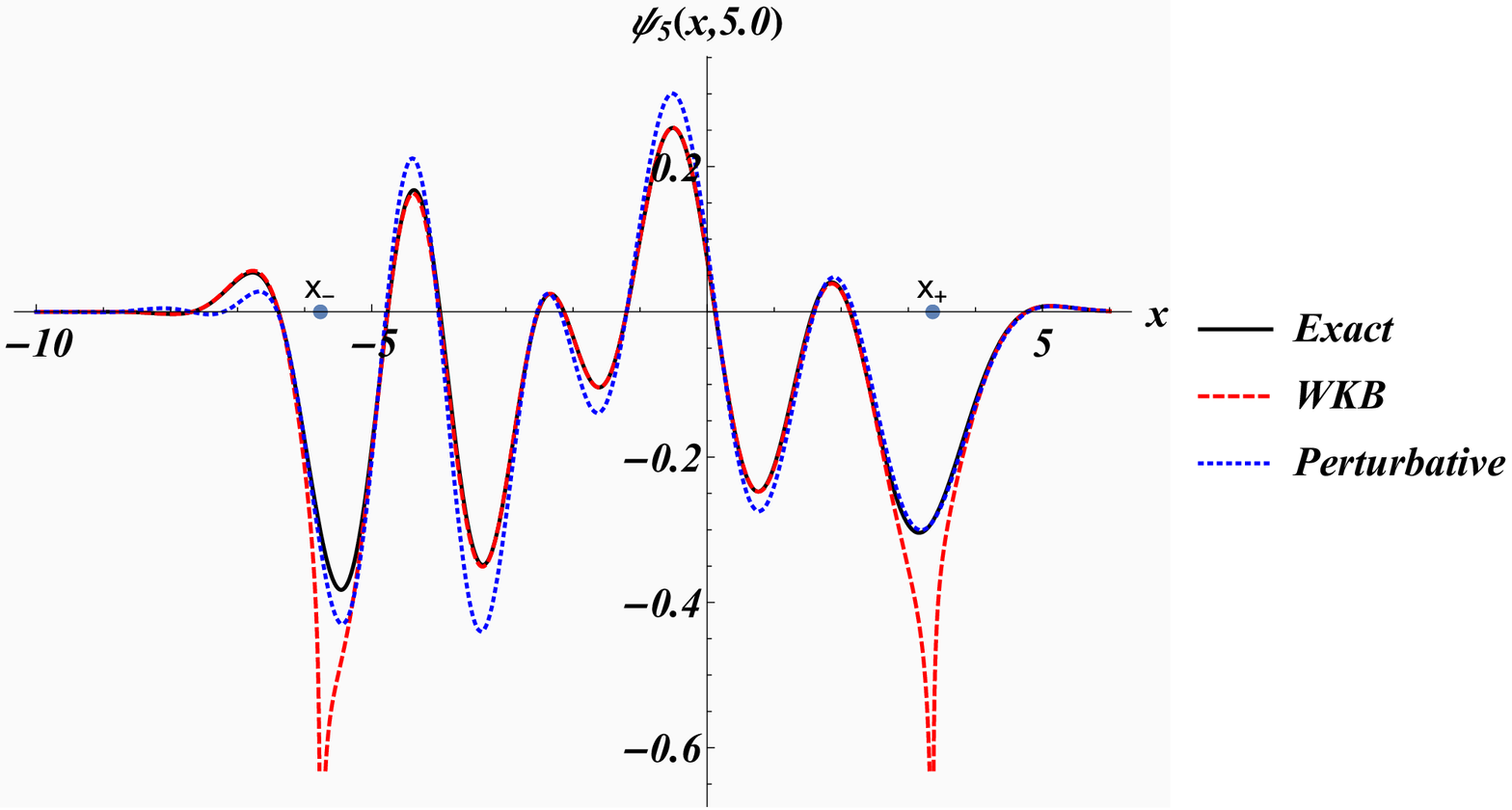, width=7.2cm}
\caption{Exact versus WKB and perturbative solutions to the 
time-dependent Schr\"{o}dinger equation (\ref{SCH}) at different times with $\hbar =1$, $E_{0}=1$, $m=1$, $n=5$, $\tau =1$, $\omega=0.5 $ and $c_{\kappa }=0.1$. The time-dependent 
classical turning points $\xi _{\pm }$ are indicated. }
        \label{Fig6}}

\section{Goldman-Krivchenko potential with time-dependent perturbation}

In trying to identify solvable systems we have seen in equation (\ref{kp})
that the value $p=-2$ is special as in that case the potential becomes the
time-independent Goldman-Krivchenko potential \cite{goldman23problems},
being a particular spiked harmonic oscillator \cite%
{hall2001generalized,znojil1992spiked}. This potential may serve also as a
benchmark for which we can solve the eigenvalue equation exactly and campare
it to the perturbative solution. Hence we take this potential as our
unperturbed system and perturb it by dropping the Stark term and replacing
it by $x^{2}E(t)$. Thus we consider the time-dependent Hamiltonian 
\begin{equation}
H(t)=\frac{p^{2}}{2m}+\frac{m\omega ^{2}}{2}x^{2}+\frac{m\Omega ^{2}}{2}%
\frac{1}{x^{2}}+x^{2}E(t).
\end{equation}%
It follows from above that the invariant for this system is 
\begin{equation}
I(t)=\frac{1}{2}\left[ \sigma ^{2}p^{2}-m\sigma \dot{\sigma}\{x,p\}+\left(
m^{2}\dot{\sigma}^{2}+\frac{\tau m^{2}+2mE_{0}}{\sigma ^{2}}\right)
x^{2}+m^{2}\sigma ^{2}\Omega ^{2}\frac{1}{x^{2}}\right] ,
\end{equation}%
with constraint $E(t)=E_{0}/\sigma ^{4}$ and $\sigma $ satisfying the
EP-equation (\ref{Erma}). Also in this case we may complete the remaining
steps in the Lewis-Riesenfeld approach and hence compare the exact and the
perturbative solution.\ We identify $E_{0}\ll 1$ as the expansion parameter
in the perturbative series.

\subsection{Testing the approximate solution}

\subsubsection{Exact computation}

Using the same similarity and variable transformation as in (\ref{Iha}) we
obtain the time-independent invariant 
\begin{equation}
\hat{I}=UIU^{-1}=-\frac{1}{2}\partial _{\xi }^{2}+\frac{1}{2}\left( \tau
m^{2}+2mE_{0}\right) \xi ^{2}+\frac{m^{2}\Omega ^{2}}{2}\frac{1}{\xi ^{2}}.
\end{equation}%
We solve the eigenvalue equation $\hat{I}\Xi (\zeta )=\lambda \Xi (\zeta )$
exactly obtaining the solution%
\begin{equation}
\Xi (\zeta )=\zeta ^{(1+b/2)}e^{-\frac{a}{2}\zeta ^{2}}\left[ c_{1}L_{\nu
_{-}}^{b/2}\left( a\zeta ^{2}\right) +c_{2}U\left( \nu _{+},1+b/2,\frac{m%
\sqrt{\tau }\zeta ^{2}}{\hbar }\right) \right] ,
\end{equation}%
where $L_{\nu }^{\mu }(z)$ denotes the generalized Laguerre polynomials, $%
U\left( \nu ,\mu ,z\right) $ the confluent hypergeometric function and $\nu
_{\pm }:=\pm \left( 2+b-2\lambda /a\right) /4$, $a=\sqrt{\tau m^{2}+2mE_{0}}$%
, $b:=\sqrt{1+4m^{2}\Omega ^{2}}$. Demanding again that the eigenfunctions
vanish asymptotically, i.e. $\lim_{\zeta \rightarrow \pm \infty }$ $\Xi
(\zeta )=0$, imposes $\nu _{\pm }=n\in \mathbb{N}_{0}$ and thus quantizes $%
\lambda $. We discard the solution related to $U$, as its corresponding
eigenvalues are not bounded from below, leading to the eigenfuctions and
eigenvalues%
\begin{equation}
\Xi _{n}(\zeta )=c_{1}\zeta ^{(1+b/2)}e^{-\frac{a}{2}\zeta
^{2}}L_{n}^{b/2}\left( a\zeta ^{2}\right) ,\qquad \lambda _{n}=a(2n+1+b/2).
\label{exl}
\end{equation}%
Assembling everything we obtain for the operator $I(t)$ the normalized
eigenfunction from $U^{-1}\Xi _{n}(x/\sigma )$ as \ 
\begin{equation}
\phi _{n}(x)=\sqrt{\frac{2n!}{[1-(-1)^{b}]\Gamma \left( 1+n+b/2\right) }}%
\left( \frac{a}{\sigma ^{2}}\right) ^{(2+b)/4}x^{(1+b)/2}e^{-\frac{a}{%
2\sigma ^{2}}x^{2}}L_{n}^{b/2}\left( \frac{ax^{2}}{\sigma ^{2}}\right) e^{im%
\dot{\sigma}x^{2}/2\sigma \hbar }.  \label{phi}
\end{equation}%
when $\func{Re}b>-2$ and $\func{Re}(a/\sigma ^{2})>0$. This completes the
second step in the Lewis-Riesenfeld approach. In the third and last step we
determine the phase $\alpha $ by means of (\ref{pht}). The right hand side
is computed once more to%
\begin{equation}
\left\langle \phi _{n}\right\vert i\partial _{t}-H(t)\left\vert \phi
_{n}\right\rangle =-\frac{\lambda _{n}}{m\sigma ^{2}}
\end{equation}%
so that the phase acquires the same form as in the previous example 
\begin{equation}
\alpha _{n}(t)=-\frac{1}{m\sqrt{\tau }}\lambda _{n}\arctan \left[ \frac{%
\sqrt{\tau }\tan \left( \omega t\right) }{\omega }\right] .
\end{equation}%
Let us now compare these expressions with those obtained in the perturbative
computation.

\subsubsection{Perturbative computation}

We treat the term $V_{p}(x,t)=x^{2}E(t)$ with $E(t)=E_{0}/\sigma ^{4}$ and $%
E_{0}\ll 1$ in the Hamiltonian as a perturbation. Accordingly we split up
the invariant (\ref{II}) as $I(t)=I_{0}(t)+E_{0}I_{p}(t)$ with 
\begin{equation}
I_{0}(t)=\frac{1}{2}\left[ \sigma ^{2}p^{2}-m\sigma \dot{\sigma}%
\{x,p\}+\left( m^{2}\dot{\sigma}^{2}+\frac{\tau m^{2}}{\sigma ^{2}}\right)
x^{2}+m^{2}\sigma ^{2}\Omega ^{2}\frac{1}{x^{2}}\right] ,\quad ~~I_{p}(t)=%
\frac{m}{\sigma ^{2}}x^{2}.
\end{equation}%
The zeroth order wavefunction $\left\vert \phi _{n}^{(0)}\right\rangle $ is
simply $\phi _{n}(x)$ in (\ref{phi}) with $E_{0}=0$. From (\ref{I0}) and (%
\ref{L1}) we compute first two terms in the perturbative series for the
eigenvalues%
\begin{eqnarray}
\lambda _{n}^{(0)} &=&\left( 2n+1+\sqrt{1+4m^{2}\Omega ^{2}}/2\right) m\sqrt{%
\tau },\quad \\
\lambda _{n}^{(1)} &=&\left\langle \phi _{n}^{(0)}\right\vert
I_{p}(t)\left\vert \phi _{n}^{(0)}\right\rangle =\left( 2n+1\right) \sqrt{%
\frac{1+4m^{2}\Omega ^{2}}{\tau }}.
\end{eqnarray}%
As expected the eigenvalues are time-independent and $\lambda
_{n}^{(0)}+E_{0}\lambda _{n}^{(1)}$ corresponds to (\ref{exl}) expanded to
first order in $E_{0}$. Next we need to compute the infinite sum in (\ref{L1}%
) to determine the corrections to the wavefunctions. In this case there are
only two terms contributing in the infinite sum. We compute%
\begin{equation}
\left\vert \phi _{n}^{(1)}\right\rangle =\frac{1}{2m\tau }\left[ \sqrt{%
(n+1)(n+1+b/2)}\left\vert \phi _{n+1}^{(0)}\right\rangle -\sqrt{n(n+b/2)}%
\left\vert \phi _{n-1}^{(0)}\right\rangle \right] .  \label{p1}
\end{equation}%
In the last step we compute the perturbed expression for the phase $\alpha
_{n}^{(1)}(t)$ using equation (\ref{pht}). Once more we find up to first
order%
\begin{equation}
\alpha _{n}^{(1)}(t)=-\frac{\lambda _{n}^{(0)}+c_{\kappa }\lambda _{n}^{(1)}%
}{m\sqrt{\tau }}\arctan \left[ \frac{\sqrt{\tau }\tan \left( \omega t\right) 
}{\omega }\right] ,  \label{p2}
\end{equation}%
so that we have obtained the full perturbative solution to the TDSE as $%
\left\vert \psi _{n}\right\rangle ^{(1)}$ as defined in (\ref{phi1}).

\subsubsection{Exact versus perturbative solutions}

As previously, we compute several physical quantities to compare the exact
and the perturbative solution. The momentum, position, squared momentum and
squared position expectation values are computed to 
\begin{equation}
\begin{array}{ll}
\left\langle \psi _{n}\right\vert x\left\vert \psi _{n}\right\rangle =0,~ & 
\left\langle \psi _{n}\right\vert x^{2}\left\vert \psi _{n}\right\rangle =%
\frac{(2n+\ell +3/2)\sigma ^{2}}{\sqrt{\tau m^{2}+2mE_{0}}}, \\ 
\left\langle \psi _{n}\right\vert p\left\vert \psi _{n}\right\rangle =0, & 
\left\langle \psi _{n}\right\vert p^{2}\left\vert \psi _{n}\right\rangle =%
\frac{(4n+2)\ell +2n+3/2}{2\ell +1}\frac{\sqrt{\tau m^{2}+2mE_{0}}}{\sigma
^{2}}+\frac{(2n+\ell +3/2)m\dot{\sigma}^{2}}{\sqrt{\tau m^{2}+2mE_{0}}}.%
\end{array}%
\end{equation}%
In order to achieve convergence we had to impose the additional constraint $%
b=2\ell +1$ with $\ell \in \mathbb{N}_{0}$. The uncertainty relation becomes%
\begin{equation}
\Delta x\Delta p=\sqrt{\frac{(2n+\ell +3/2)[(4n+2)\ell +2n+3/2]}{2\ell +1}+%
\frac{m(2n+\ell +3/2)^{2}\sigma ^{2}\dot{\sigma}^{2}}{(\tau m+2E_{0})^{2}}},
\end{equation}%
with the lower bound $\Delta x\Delta p\geq 1/2$ always well respected.

Using the perturbed solutions (\ref{p1}) and (\ref{p2}) we compute%
\begin{equation}
\begin{array}{l}
\left\langle \psi _{n}\right\vert x\left\vert \psi _{n}\right\rangle
^{(1)}=0,~~\left\langle \psi _{n}\right\vert x^{2}\left\vert \psi
_{n}\right\rangle ^{(1)}=(2n+\ell +3/2)(m\tau -E_{0})\frac{\sigma ^{2}}{\tau
m^{2}}, \\ 
\left\langle \psi _{n}\right\vert p\left\vert \psi _{n}\right\rangle
^{(1)}=0,~~\left\langle \psi _{n}\right\vert p^{2}\left\vert \psi
_{n}\right\rangle ^{(1)}=\frac{(4n+2)\ell +2n+3/2}{2\ell +1}\frac{(m\tau
+E_{0})}{\sqrt{\tau }\sigma ^{2}}+\frac{(2n+\ell +3/2)(\tau m-E_{0})\dot{%
\sigma}^{2}}{\tau ^{3/2}}.%
\end{array}%
\end{equation}%
The approximated uncertainty relation results to%
\begin{equation}
\Delta x\Delta p^{(1)}=\sqrt{\frac{(2n+\ell +3/2)[(4n+2)\ell +2n+3/2]}{2\ell
+1}(1-\frac{E_{0}^{2}}{m^{2}\tau ^{2}})}.
\end{equation}%
We are now in the position to compare the exact and the pertubative
solution. In figure \ref{Fig3} we compare the time-dependent expectation
values for $x^{2}$ and $p^{2}$ computed in an exact way with those computed
in a perturbative fashion. As in the previous example, the agreement is very
good for small values of the expansion parameter, $E_{0}$ in this case.
Overall the agreement is increasing for large values of $n$ as well as $m$
and for $\omega $ approaching $\sqrt{\tau }$.

\FIGURE{ \epsfig{file=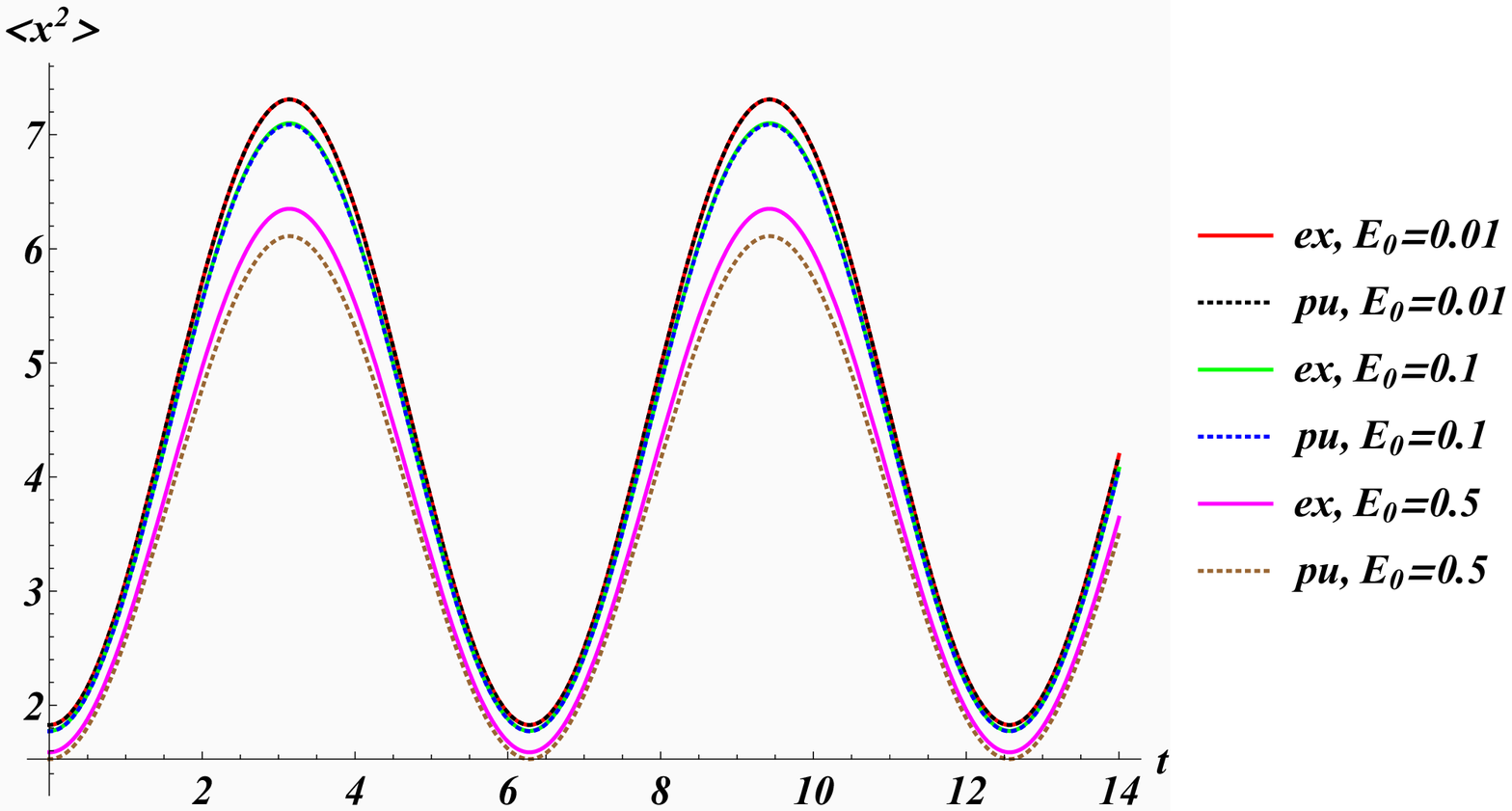,width=7.2cm} \epsfig{file=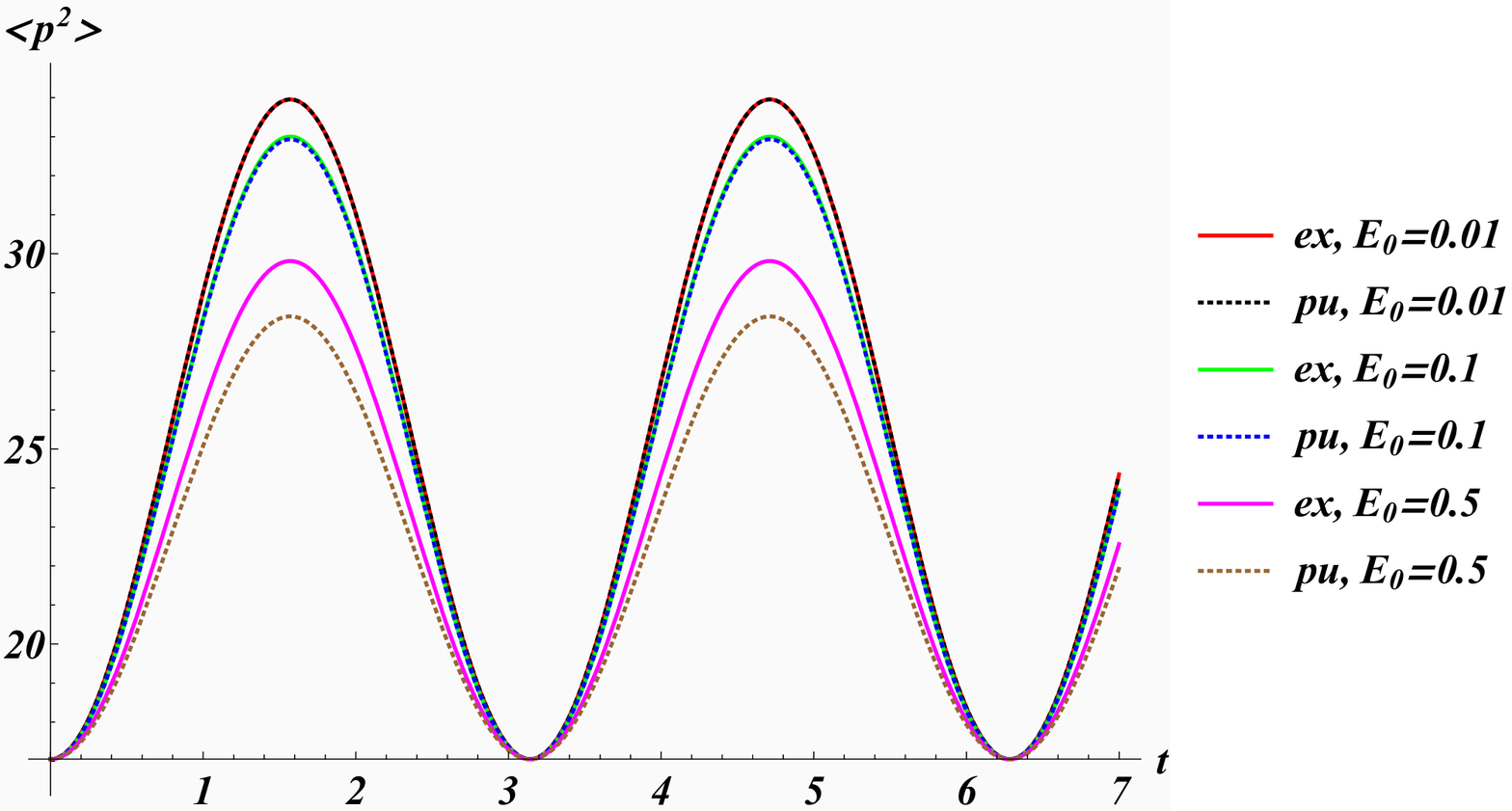,width=7.2cm} 
        \caption{Exact versus perturbative expectation values for  $x^{2}$, $p^2$, for $\omega =1/2$, $\tau =1$, $\ell=2$, $m=3$ and $n=1$ for different values of the expansion parameter $E_0$.}
        \label{Fig3}}

As in the previous example we also compute the autocorrelation function (\ref%
{auto}) as it captures well the effect from the time-dependent phase $\alpha
(t)$. We depict this function in figure \ref{Fig4}. Once more, the overall
agreement decreases for larger values of $n$.

\FIGURE{ \epsfig{file=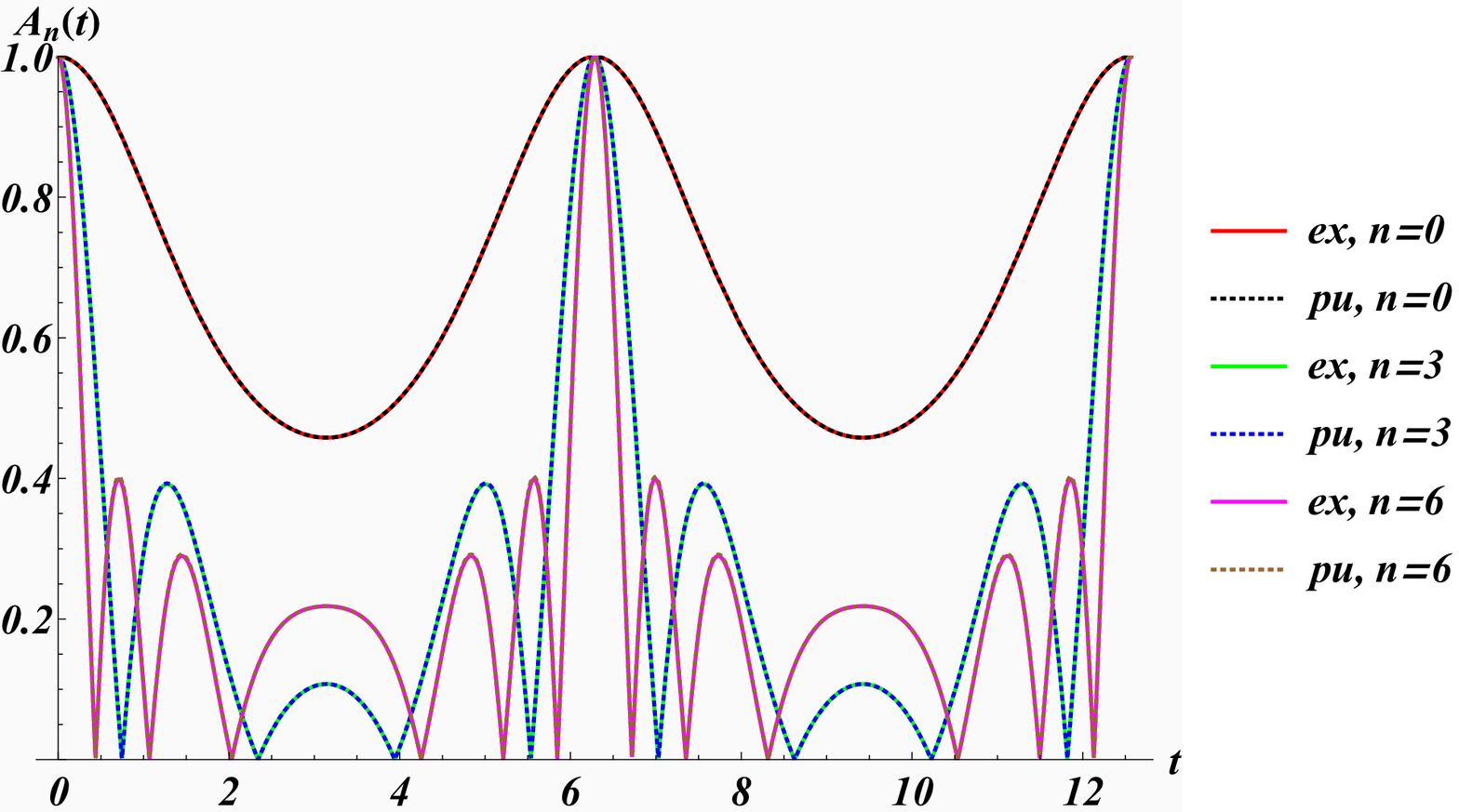,width=7.2cm} \epsfig{file=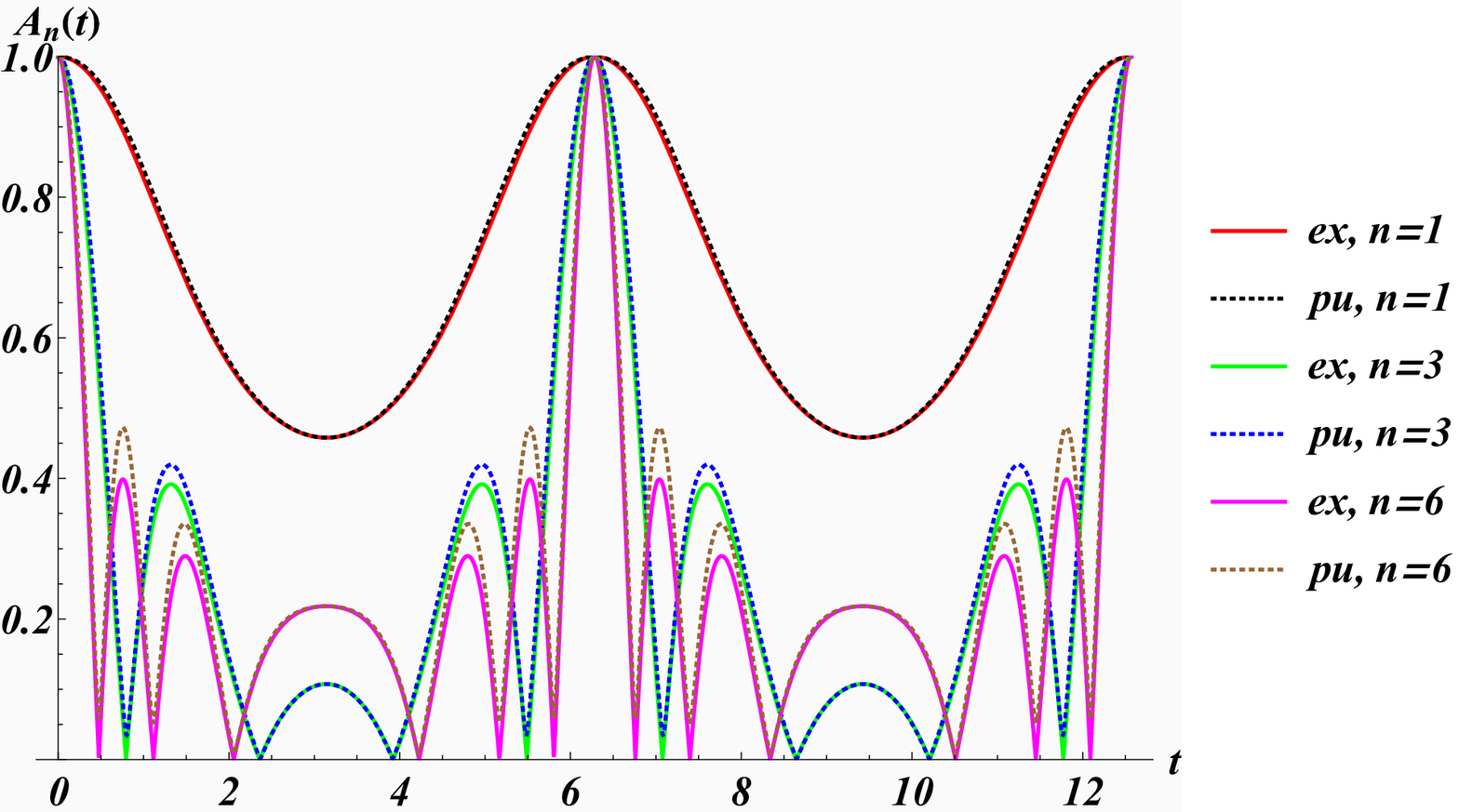,width=7.2cm}
        \caption{Exact versus perturbative autocorrelation function for $\omega =1/2$, $\tau =1$, $m=3$, $\ell=2$ and different values 
         for $n$ with $E_0=0.1$ in the left panel and $E_0=0.5$ in the right panel.}
        \label{Fig4}}

\section{Conclusions}

We have explored the possibilty of a modified approximated Lewis and
Riesenfeld method by solving the time-independent eigenvalue equation in the
second step by means of standard time-independent perturbation theory. We
have tested the quality of this approach for two classes of optical
potentials by comparing the exact solutions obtained from the completely
exact solution of the Lewis-Riesenfeld approach to the approximated ones,
the perturbative approach and the WKB approximation. We computed some
standard expectation values and the autocorrelation functions in two
alternative ways. For the pertubative approach we found in general good
agreement which is naturally improved in quality for smaller values of the
expansion parameters. The WKB approximation is not limited to these small
parameters and only deviates significantly at the turning points. Our
semi-exactly solvable approach significantly widens the scope of the
Lewis-Riesenfeld method and allows to tackle more complicated physical
situations that are not possible to treat when insisting on full exact
solvabilty

While in this paper we were mainly concerned with a proof of concept related
to the modification of the Lewis-Riesenfeld method, it would naturally be
very interesting now to apply the scheme to study more interesting and
challenging physical phenomena. To this end one may carry out further
approximations. Here we are still assuming that the first step in the
approach, i.e. the computation of the invariant, can be carried out in an
exact manner. Naturally one may also weaken this requirement and work with
an approximated invariant in the second step. We will present this
possibility elsewhere \cite{AFRTprep}.

\textbf{Acknowledgments:} RT is supported by a City, University of London
Research Fellowship. We thank Carla Figueira de Morisson Faria for
discussions and comments.

\newif\ifabfull\abfulltrue


\end{document}